\documentclass[a4paper]{article}

\usepackage[english]{babel}
\usepackage[utf8x]{inputenc}
\usepackage[T1]{fontenc}

\usepackage[a4paper,top=3cm,bottom=2cm,left=3cm,right=3cm,marginparwidth=1.75cm]{geometry}
\usepackage{setspace}
\usepackage[table]{xcolor}
\usepackage[title]{appendix}
\usepackage{amsmath}
\usepackage{mathrsfs}
\usepackage{amsthm}
\usepackage{bbm}

\usepackage{amssymb, multirow}
\usepackage{enumitem}
\usepackage{graphicx}
\usepackage[colorinlistoftodos]{todonotes}
\usepackage[colorlinks=true, allcolors=blue]{hyperref}
\usepackage{wrapfig}
\usepackage{subcaption}
\usepackage{booktabs,siunitx}
\usepackage{natbib}
\usepackage[linesnumbered,boxruled,lined]{algorithm2e}
\usepackage{algpseudocode}

\usepackage{url}

\title{An efficient EM algorithm for both element-wise and structural missingness in matrix-variate normal mixture models}

\date{}
\author{Hanzhang Lu\footnote{Corresponding Author: University of British Columbia Okanagan Campus, Kelowna, BC, Canada, V1V 1V7. Email: hanzhang.lu@ubc.ca}, Jeffrey L.\ Andrews, Ryan P.\ Browne }

\begin{document}
\maketitle

\begin{abstract}
Matrix-variate data with missing entries arise frequently in applications where observations are naturally organized as two-dimensional arrays. Although the matrix normal distribution provides a parsimonious model through its Kronecker covariance structure, standard EM estimation can be computationally expensive because arbitrary missingness patterns typically destroy this separability in the E-step. In this paper, we propose an efficient partial EM algorithm for matrix-variate normal data with missing entries. The proposed method updates the conditional mean and covariance of the missing component through coordinate-wise approximations, avoiding repeated inversion of pattern-specific covariance matrices and avoiding construction of the full vectorized covariance matrix. We further develop a specialized update for submatrix missingness, where the missing-block precision retains a Kronecker product structure, and the covariance update can be carried out independently in the row and column directions. Simulation studies show that the proposed methods substantially reduce computation time compared with exact EM while preserving nearly identical observed-data likelihood across a range of dimensions and missing proportions. A real-data application to hyperspectral image patches demonstrates that the proposed imputation strategy can be embedded within a matrix-variate mixture model for simultaneous imputation and clustering.
     
\end{abstract} 

\section{Introduction} \label{sec:intro}
Matrix-valued data are prevalent in contemporary applications such as spatiotemporal measurements, image patches, multichannel sensor recordings, and biological assays. In these contexts, each observation constitutes a two-dimensional array, with rows and columns that often represent distinct scientific entities. Missing entries frequently occur in such datasets due to sensor failures, occluded images, or incomplete experimental assays. Consequently, effective imputation methods should leverage dependencies along both dimensions of the matrix, rather than reducing the data to an unstructured vector \citep{little2019statistical, schafer1997analysis}.

The matrix-variate normal (MVN) distribution \citep{dawid1981some, gupta2018matrix} serves as the matrix-variate analog of the normal distribution. It provides a fundamental framework for modeling matrix-variate data, offering mathematical tractability and the ability to capture dependencies across both rows and columns. Consider a $p \times q$ random matrix $\mathcal{Y}$. The matrix $\mathcal{Y}$ is said to follow an MVN distribution, denoted as $\mathcal{Y} \sim \mathcal{N}_{p \times q}(\mathbf{M},\boldsymbol{\Sigma}_1,\boldsymbol{\Sigma}_2)$, if its probability density function is given by
\begin{equation*}
    \phi_{p \times q}(\mathbf{Y} \mid \mathbf{M},\boldsymbol{\Sigma}_1,\boldsymbol{\Sigma}_2) = \frac{\operatorname{exp}\left\{-\frac{1}{2}\operatorname{tr}\left(\boldsymbol{\Sigma}_1^{-1}(\mathbf{Y}-\mathbf{M})\boldsymbol{\Sigma}_2^{-1}(\mathbf{Y}-\mathbf{M})^\prime\right)\right\}}{(2\pi)^{\frac{pq}{2}}|\boldsymbol{\Sigma}_1|^{\frac{q}{2}}|\boldsymbol{\Sigma}_2|^{\frac{p}{2}}},
\end{equation*}
where $\mathbf{M}$ is the $p \times q$ location matrix, $\boldsymbol{\Sigma}_1$ is the $p \times p$ row covariance matrix, and $\boldsymbol{\Sigma}_2$ is the $q \times q$ column covariance matrix. Equivalently,
\begin{equation} \label{eq:mvneq}
    \operatorname{vec}(\mathcal{Y}) \sim \mathcal{N}_{pq}(\operatorname{vec}(\mathbf{M}), \boldsymbol{\Sigma}_2\otimes\boldsymbol{\Sigma}_1),
\end{equation}
where $\mathcal{N}_{pq}(\operatorname{vec}(\mathbf{M}), \boldsymbol{\Sigma}_2\otimes\boldsymbol{\Sigma}_1)$ denotes the $pq$-dimensional multivariate normal distribution with mean vector $\operatorname{vec}(\mathbf{M})$ and covariance matrix $\boldsymbol{\Sigma}_2\otimes\boldsymbol{\Sigma}_1$. 
The Kronecker covariance structure reduces the number of free covariance parameters from
$pq(pq+1)/2$ for an unrestricted covariance matrix to
$p(p+1)/2+q(q+1)/2-1$, where the subtraction of one accounts for the scale non-identifiability of the two covariance factors.
Consequently, the matrix normal model is particularly suitable for high-dimensional matrix-valued observations where an unrestricted multivariate normal covariance is statistically unstable or computationally infeasible \citep{dutilleul1999mle,allen2010transposable}.

Missing-value imputation within this model can be addressed using maximum likelihood estimation (MLE) via the expectation-maximization (EM) algorithm \citep{dempster77}. The E-step computes the conditional distribution of the missing entries given the observed data under the current parameters. The M-step then updates the mean and covariance factors using the corresponding conditional sufficient statistics. For complete data, estimation of the covariance factors is already coupled and is typically performed using a flip-flop or block-coordinate procedure \citep{dutilleul1999mle}. In the presence of missing data, \cite{glanz2018expectation} introduced an EM algorithm holding the same basic structure, while \cite{lachos2025algorithm} proposed an expectation-conditional maximization (ECM) algorithm \citep[ECM;][]{meng1993maximum} designed for both matrix-variate interval-censored and missing data. Both developments focus on the M-step for estimating the Kronecker covariance.

The EM algorithm is recognized for its significant computational demands, a challenge that persists in the matrix-variate setting. The primary difficulty arises because the E-step often requires inverting the covariance matrix of the observed component for each unique missingness pattern. \cite{browne2022partial} introduced the partial EM (PEM) algorithm, which updates the sufficient statistics in a coordinate-wise manner. In the matrix-variate context, arbitrary missingness patterns typically disrupt the Kronecker structure of the covariance matrix when restricted to observed entries. While the full covariance matrix is separable, the observed submatrix defined by an arbitrary mask is not necessarily representable as a smaller Kronecker product. As a result, the E-step may still necessitate vectorizing matrices and performing dense Gaussian conditioning for each missingness pattern. This process can be computationally intensive in terms of both time and memory, particularly when matrices are large, the number of observations is substantial, or missingness patterns differ across samples. In these scenarios, although the standard EM algorithm may offer statistical advantages, it often remains computationally infeasible.

Building upon the PEM framework, this work introduces a PEM for the matrix normal distribution (MPEM) that updates the conditional mean and covariance of missing entries directly in matrix form. By utilizing Kronecker precision factors, the proposed procedure avoids both pattern-specific matrix inversions and the construction of the full $pq \times pq$ covariance matrix. We further derive a specialized update for submatrix missingness, in which the precision matrix associated with the missing block maintains an exact Kronecker representation. Submatrix missingness arises naturally in applications where observations are indexed along two meaningful dimensions. For instance, localized corruption in image or video data may remove an entire spatial region, while in longitudinal or panel data, measurements may be unavailable for a subset of units over a common time interval, resulting in a missing submatrix \citep{athey2021matrix}. For fixed model parameters, the exact conditional moments correspond to fixed points of the proposed updates. Performing only a limited number of coordinate sweeps yields a computationally efficient partial E-step. The procedure is further extended to mixtures of matrix-variate normal distributions, enabling simultaneous imputation and clustering.

The remainder of this paper is organized as follows. 
Section~\ref{sec:preliminaries} reviews the EM algorithm for incomplete multivariate and matrix-variate normal data, the partial EM framework, and the scale-identifiability issue associated with Kronecker-structured covariance matrices. 
Section~\ref{sec:method} develops the proposed partial E-step for matrix-variate normal data, derives coordinate-wise updates for the conditional mean and covariance of the missing entries, presents the corresponding M-step, introduces a specialized procedure for submatrix missingness, and reports simulation results for the single-component setting. 
Section~\ref{sec:mixtures} extends the proposed method to mixtures of matrix-variate normal distributions, describes the mixture of spatial factor analyzers used in the application, and presents simulation results for the mixture setting. 
Section~\ref{sec:apps} applies the proposed methods to the Salinas Valley hyperspectral image data. 
Finally, Section~\ref{sec:summary} summarizes the main findings and concludes the paper.

\section{Preliminaries} \label{sec:preliminaries}
Prior to detailing the proposed methodology, we introduce the necessary notation and outline the foundational background relevant to this study.

\subsection{EM algorithm for imputation} \label{sec:em}
Consider independent random vectors $\mathbf{y}_1,\ldots,\mathbf{y}_N \sim \mathcal{N}_p(\boldsymbol{\mu},\boldsymbol{\Sigma}),$ where some entries of each $\mathbf{y}_i$ may be missing. Let $\mathbf{x}_i$ denote the observed component and $\mathbf{z}_i$ denote the missing component. 
The EM algorithm maximizes the likelihood by iteratively replacing the unavailable sufficient statistics of $\mathbf{z}_i$ with their conditional expectations given the observed $\mathbf{x}_i$. With a specific permutation, the mean vector and covariance matrix are partitioned according to the indices corresponding to $\mathbf{x}_i$ and $\mathbf{z}_i$ as
\begin{equation*}
    \boldsymbol{\mu} = \begin{bmatrix} \boldsymbol{\mu}_{x,i} \\ \boldsymbol{\mu}_{z,i} \end{bmatrix}, \qquad \boldsymbol{\Sigma} = \begin{bmatrix} \boldsymbol{\Sigma}_{xx,i} & \boldsymbol{\Sigma}_{xz,i} \\ \boldsymbol{\Sigma}_{zx,i} & \boldsymbol{\Sigma}_{zz,i} \end{bmatrix}.
\end{equation*}
Note that the partition may differ across observations because each observation can have a different missingness pattern. 

In the E-step, the conditional expectation and conditional covariance of the missing data are computed given the observed data $\mathbf{x}_i$ and the current parameter estimates, which are given by
\begin{equation}\label{eq:estep}
    \hat{\mathbf{z}}_i  
    = \boldsymbol{\mu}_{z\mid x,i} 
    = \boldsymbol{\mu}_{z,i} 
    + \boldsymbol{\Sigma}_{zx,i} \boldsymbol{\Sigma}_{xx,i}^{-1} 
    (\mathbf{x}_i - \boldsymbol{\mu}_{x,i}), \quad
    \hat{\mathbf{Z}}_{i} 
    = \boldsymbol{\Sigma}_{z\mid x,i} = \boldsymbol{\Sigma}_{zz,i} 
    - \boldsymbol{\Sigma}_{zx,i} 
    \boldsymbol{\Sigma}_{xx,i}^{-1} 
    \boldsymbol{\Sigma}_{xz,i}.
\end{equation}
In \eqref{eq:estep}, $\boldsymbol{\Sigma}_{xx,i}^{-1}$ depends on the observed set of observation $i$. When the data contain many distinct missingness patterns, it requires many different matrix inversions at every iteration. This missing pattern-specific inversion calculation, which is computationally expensive, is one of the main sources of the complexity. Let $\hat{\mathbf{y}}_i$ represents the the full vectors with the missing entries replaced by $\hat{\mathbf{z}}_i$, and $\hat{\mathbf{V}}_i$ represent the correction matrix with entries corresponding to missing component filled with $\hat{\mathbf{Z}}_i$. In the M-step, the mean and covariance are updated by maximizing the conditional expected log-likelihood via
\begin{equation*}
    \hat{\boldsymbol{\mu}} = \frac{1}{N} \sum_{i=1}^N \hat{\mathbf{y}}_i \ , \qquad \hat{\boldsymbol{\Sigma}} = \frac{1}{N} \sum_{i=1}^N \left[ (\hat{\mathbf{y}}_i - \hat{\boldsymbol{\mu}})(\hat{\mathbf{y}}_i - \hat{\boldsymbol{\mu}})^\prime + \hat{\mathbf{V}}_i \right].
\end{equation*}

Another important property around \eqref{eq:estep}, which will be used later, is the relation between the Schur complement of the matrix $\boldsymbol{\Sigma}$ and the precision matrix. Let $\boldsymbol{\Xi}$ represent the precision matrix. Given the partition corresponding to $\mathbf{x}_i$ and $\mathbf{z}_i$, we have
\begin{equation*}
    \boldsymbol{\Sigma} = \begin{bmatrix} \boldsymbol{\Sigma}_{xx} & \boldsymbol{\Sigma}_{xz} \\ \boldsymbol{\Sigma}_{zx} & \boldsymbol{\Sigma}_{zz} \end{bmatrix} \quad \text{and} \quad \boldsymbol{\Sigma}^{-1} = \boldsymbol{\Xi} = \begin{bmatrix} \boldsymbol{\Xi}_{xx} & \boldsymbol{\Xi}_{xz} \\ \boldsymbol{\Xi}_{zx} & \boldsymbol{\Xi}_{zz} \end{bmatrix}.
\end{equation*}
The property can be written as
\begin{equation} \label{eq:Schur}
    \boldsymbol{\Sigma}_{z\mid x}^{-1} = \boldsymbol{\Xi}_{zz}, \quad \text{and } \quad \boldsymbol{\Sigma}_{zx}\boldsymbol{\Sigma}^{-1}_{xx}=-\boldsymbol{\Xi}^{-1}_{zz}\boldsymbol{\Xi}_{zx}.
\end{equation}

\subsection{EM algorithm for matrix normal distribution} \label{sec:emformat}
In the matrix-variate setting, according to \eqref{eq:mvneq}, the main difference concerns the Kronecker product covariance compared to the multivariate case. As mentioned in Section~\ref{sec:intro}, since the Kronecker structure is lost under arbitrary missingness, the E-step is performed on the vectorized matrices in the same way as in the multivariate case. Therefore, the difficulty lies in the Kronecker covariance.
In the M-step, to estimate the Kronecker covariance, \cite{glanz2018expectation} project the conditional covariance onto the row and column covariance spaces.
Consider independent random matrices $\mathcal{Y}_1,\ldots,\mathcal{Y}_N \sim \mathcal{N}_{p\times q}(\mathbf{M},\boldsymbol{\Sigma}_1,\boldsymbol{\Sigma}_2),$ where entries of each $\mathcal{Y}_i$ are missing at random. Let $\mathbf{y}_i$ be the vectorization of $\mathcal{Y}_i$, $\boldsymbol{\mu}$ be the vectorization of $\mathbf{M}$, and $\boldsymbol{\Sigma} = \boldsymbol{\Sigma}_2 \otimes \boldsymbol{\Sigma}_1$. Consequently, in the E-step, with the same notation, the same calculation as \eqref{eq:estep} is conducted to gain the sufficient statistics $\hat{\mathbf{z}}_i$ and $\hat{\mathbf{Z}}_i$. 

To estimate $\boldsymbol{\Sigma}_1$ and $\boldsymbol{\Sigma}_2$ within the Kronecker structure, the full correction matrix $\hat{\mathbf{V}}_i$ is projected onto the row and column covariance spaces by taking the element-wise partial derivative of the marginal log-likelihood. Let $\boldsymbol{\Xi}_1 = \boldsymbol{\Sigma}_1^{-1}$, $\boldsymbol{\Xi}_2 = \boldsymbol{\Sigma}_2^{-1}$, and $\hat{\mathbf{R}}_i$ and $\hat{\mathbf{C}}_i$ denote the projected terms in row and column covariance spaces, respectively. Taking the column covariance as an example, the $(k,l)$ entry of $\mathbf{C}_i$ is calculated as
\begin{equation*}
    \left[\hat{\mathbf{C}}_i\right]_{(k,l)}= \operatorname{tr} \left[ \frac{\partial}{\partial (\boldsymbol{\Xi}_2)_{kl}} \{ (\boldsymbol{\Xi}_2 \otimes \boldsymbol{\Xi}_1) \hat{\mathbf{V}}_i \} \right] = \operatorname{tr} \left[ \left\{ \frac{\partial \boldsymbol{\Xi}_2}{{\partial (\boldsymbol{\Xi}_2)_{kl}}} \otimes \boldsymbol{\Xi}_1 \right\} \hat{\mathbf{V}}_i \right]. 
\end{equation*}
Similarly, an element-wise update equation for $\hat{\mathbf{R}}_i$ can also be derived through the partial derivative. With all the projected terms $\hat{\mathbf{R}}_i$ and $\hat{\mathbf{C}}_i$, in the M-step, for the row covariance matrix, the update equation is
\begin{equation*}
    \hat{\boldsymbol{\Sigma}}_1 = \frac{1}{Nq} \sum_{i=1}^N \left[\left(\mathbf{Y}_i-\mathbf{M}\right) \hat{\boldsymbol{\Xi}}_2\left(\mathbf{Y}_i-\mathbf{M}\right)^\prime + \hat{\mathbf{R}}_i\right],
\end{equation*}
and for the column covariance matrix, the update equation is
\begin{equation*}
    \hat{\boldsymbol{\Sigma}}_2 = \frac{1}{Np} \sum_{i=1}^N \left[\left(\mathbf{Y}_i-\mathbf{M}\right) ^\prime\hat{\boldsymbol{\Xi}}_1\left(\mathbf{Y}_i-\mathbf{M}\right) + \hat{\mathbf{C}}_i\right].
\end{equation*}

With the conditional mean $\hat{\mathbf{Y}}_i$ and covariance $\hat{\mathbf{V}}_i$, \cite{lachos2025algorithm} also provide an M-step by taking the Cholesky factorization of the expected complete-data scatter matrix
\begin{equation*}
    \hat{\boldsymbol{\Delta}}_i = \operatorname{vec}(\mathbf{Y}_i-\hat{\mathbf{M}})\operatorname{vec}(\mathbf{Y}_i-\hat{\mathbf{M}})^\prime + \hat{\mathbf{V}}_i,
\end{equation*}
and reshapes the columns of the Cholesky factor into $p\times q$ matrices. These matrices are then used to conduct the flip-flop M-step for estimating the row covariance $\boldsymbol{\Sigma}_1$ and the column covariance $\boldsymbol{\Sigma}_2$.


\subsection{Partial EM algorithm} \label{sec:pem} 
Since the E-step is one of the main sources of computational complexity, \cite{browne2022partial} propose a partial E-step that approximates the sufficient statistics at each iteration. Following \cite{neal1998view}, PEM views the EM algorithm as minimizing the Kullback--Leibler (KL) divergence between the distribution of the missing component and the true conditional distribution. But instead of minimizing the KL divergence exactly, PEM performs a partial E-step that only reduces the KL divergence to estimate $\hat{\mathbf{z}}_i$ and $\hat{\mathbf{Z}}_i$. For the missing-data mean update, \cite{browne2022partial} rewrite the problem as minimizing a full quadratic form
\begin{equation*}
    J(\hat{\mathbf{y}})=\left(\hat{\mathbf{y}}-\boldsymbol{\mu}\right)^\prime \boldsymbol{\Sigma}^{-1}\left(\hat{\mathbf{y}}-\boldsymbol{\mu}\right). 
\end{equation*}
By working with the precision matrix $\boldsymbol{\Xi} = \boldsymbol{\Sigma}^{-1}$ and coordinate descent, for a missing coordinate $j$ of observation $i$, PEM updates the current completed value by the coordinate-wise conditional mean
\begin{equation*}
    \hat{y}_{ij} =
    \mu_j
    - \frac{1}{\xi_{jj}}
    \boldsymbol{\xi}_{j,-j}
    \left(
        \hat{\mathbf y}_{i,-j}
        - \boldsymbol{\mu}_{-j}
    \right),
\end{equation*}
where $\xi_{jj}$ is the $j$th diagonal element of
$\boldsymbol{\Xi}$ and $\boldsymbol{\xi}_{j,-j}$ is the $j$th row with the $j$th entry removed. Observed coordinates are kept fixed at their observed
values. For the missing-data covariance update, it shows that minimizing the KL divergence is equivalent to optimizing another convex function
\begin{equation*}
    J(\hat{\mathbf{V}})=\operatorname{tr}\{(\boldsymbol{\Sigma}-\hat{\mathbf{V}})\boldsymbol{\Sigma}^{-1}(\boldsymbol{\Sigma}-\hat{\mathbf{V}})\}.
\end{equation*}
Again, via coordinate descent, analogous row-and-column updates are applied to
$\hat{\mathbf V}_i$, if $j$ corresponds to the missing component, which can be given by
\begin{equation*}
    \hat{\mathbf V}_{i,j} = \boldsymbol{\Sigma}_{j} -
\left(
    \boldsymbol{\sigma}_{j,-j}
    -
    \hat{\mathbf v}_{i,-j}
\right)^\prime
\boldsymbol{\Sigma}_{-j,-j}^{-1}
\left(
    \boldsymbol{\Sigma}_{-j}
    -
    \hat{\mathbf V}_{i,-j,\cdot}
\right),
\end{equation*}
where $\hat{\mathbf V}_{i,j}$ is the $j$th row of $\hat{\mathbf V}_i$, $\hat{\mathbf V}_{i,-j}$ is the matrix obtained from $\hat{\mathbf V}_i$ by deleting row $j$, $\boldsymbol{\Sigma}_{j}$ is the $j$th row of $\boldsymbol{\Sigma}$, $\boldsymbol{\Sigma}_{-j}$ is the matrix obtained from $\boldsymbol{\Sigma}$ by deleting row $j$, $\boldsymbol{\Sigma}_{-j,-j}$ is the principal submatrix of $\boldsymbol{\Sigma}$ obtained by deleting row and column $j$, $\boldsymbol{\sigma}_{j,-j}$ is the $j$th column of $\boldsymbol{\Sigma}$ with the $j$th entry removed, and $\hat{\mathbf v}_{i,-j}$ is the $j$th column of $\hat{\mathbf V}_i$ with the $j$th entry removed. For the inverse $\boldsymbol{\Sigma}_{-j,-j}^{-1}$, it can be avoided via
\begin{equation*}
    \boldsymbol{\Sigma}_{-j,-j}^{-1} = \left[\mathbf{I}_{p-1}+\frac{1}{1-\boldsymbol{\xi}^\prime_{j,-j}\boldsymbol{\sigma}_{j,-j}}\boldsymbol{\xi}_{j,-j}\boldsymbol{\sigma}^\prime_{j,-j}\right]\boldsymbol{\Xi}_{-j,-j},
\end{equation*}
when $\boldsymbol{\xi}^\prime_{j,-j}\boldsymbol{\sigma}_{j,-j} \ne 1$, $\boldsymbol{\Sigma}_{-j,-j}$ is the principal submatrix of $\boldsymbol{\Sigma}$ with row and column $j$ removed, and $\boldsymbol{\Xi}_{-j,-j}$ is the principal submatrix of $\boldsymbol{\Xi}$ with row and column $j$ deleted. Then the $j$th column of $\hat{\mathbf V}_i$ is set to be equal to its $j$th row. The blocks corresponding to the observed component are set to zero matrices. Note that in the EM algorithm, evaluating the likelihood requires the precision matrix $\boldsymbol{\Xi}$ at every iteration. If $\boldsymbol{\Xi}$ is assumed to be known, the partial E-step avoids the calculation of the matrix inverse corresponding to the unique missingness patterns.

\subsection{Identifiability} \label{sec:identifiability}
A well-known issue in Kronecker-structured covariance models is the scale non-identifiability of the two covariance factors. Without additional constraints, the two Kronecker factors are not uniquely determined, since for any constant $\alpha\ne 0$, $\boldsymbol{\Sigma}_2 \otimes \boldsymbol{\Sigma}_1 = \frac{1}{\alpha}\boldsymbol{\Sigma}_2 \otimes \alpha\boldsymbol{\Sigma}_1 $. Thus, different pairs of covariance factors can generate the same overall covariance matrix. Following \cite{sharp2023dual}, we resolve this non-identifiability by introducing a variance scalar $\sigma^2$ and imposing the constraints $|\boldsymbol{\Sigma}_1|=1$ and $|\boldsymbol{\Sigma}_2|=1$. Now, the current full covariance matrix becomes $\boldsymbol{\Sigma} =\sigma^2 \boldsymbol{\Sigma}_2 \otimes \boldsymbol{\Sigma}_1$. And the corresponding precision matrix becomes $\boldsymbol{\Xi}=\frac{1}{\sigma^2}\boldsymbol{\Xi}_2\otimes\boldsymbol{\Xi}_1$.

\section{Methods for missingness in matrix-variate normals} \label{sec:method}
As mentioned in Section~\ref{sec:emformat}, since arbitrary missingness often disrupts the Kronecker structure, the E-step cannot utilize the separability of the covariance to compute the sufficient statistics. If the dimensionality is high, performing \eqref{eq:estep} may require substantial computational and memory resources. However, inspired by PEM, we can preserve the Kronecker structure during Gaussian conditioning and reduce the computational and memory burden if the conditional moments are updated coordinate-wise. Here, the problem is also considered starting from the KL divergence. Due to the arbitrary missing patterns, the observed and missing components may not be the matrices, so here we start with vectorization. Again, with $\mathcal{Y} \sim \mathcal{N}_{p\times q}(\mathbf{M},\boldsymbol{\Sigma}_1,\boldsymbol{\Sigma}_2,\sigma^2),$ $\operatorname{vec}(\mathcal{Y})\sim \mathcal{N}_{pq}(\boldsymbol{\mu},\boldsymbol{\Sigma})$, where $\boldsymbol{\mu}$ is the vectorization of $\mathbf{M}$, and $\boldsymbol{\Sigma} = \sigma^2\boldsymbol{\Sigma}_2 \otimes \boldsymbol{\Sigma}_1$. The KL divergence can be written as
\begin{equation} \label{eq:KL}
    \mathcal{D}_{\text{KL}} = \frac{1}{2} \left[ (\hat{\mathbf{z}} - \boldsymbol{\mu}_{z|x})^\prime \boldsymbol{\Sigma}_{z|x}^{-1} (\hat{\mathbf{z}} - \boldsymbol{\mu}_{z|x})+ \operatorname{tr}\{ \boldsymbol{\Sigma}_{z|x}^{-1} \hat{\mathbf{Z}} \} - \ln \left( \frac{|\hat{\mathbf{Z}}|}{|\boldsymbol{\Sigma}_{z|x}|} \right) \right],
\end{equation}
where $\boldsymbol{\mu}_{z|x}$ is the conditional mean of the missing component, and $\boldsymbol{\Sigma}_{z|x}$ is the conditional covariance matrix. Because of the decoupling property of the KL divergence, it is optimized with respect to the mean and covariance separately. Hereafter, all updates and derivations are assumed to apply to an arbitrary matrix, with observation subscripts omitted.

\subsection{E-step for conditional mean}
We first consider the part of the KL-divergence objective that depends on the conditional mean. Let $\hat{\mathbf{Y}}$ denote the current completed data matrix, in which the observed entries are fixed at their observed values and the missing entries are replaced by their current estimates. Up to terms that do not depend on $\hat{\mathbf{Y}}$, the relevant objective can be written in matrix form as
\begin{equation}
    J(\hat{\mathbf{Y}})=\frac{1}{2} \operatorname{tr}\left[\boldsymbol{\Xi}_1\left(\hat{\mathbf{Y}}-{\mathbf{M}}\right)\boldsymbol{\Xi}_2\left(\hat{\mathbf{Y}}-{\mathbf{M}}\right)^\prime\right],
\end{equation}
where $\boldsymbol{\Xi}_1=\boldsymbol{\Sigma}_1^{-1}$ and $\boldsymbol{\Xi}_2=\boldsymbol{\Sigma}_2^{-1}$ are the row and column precision matrices, respectively. This representation is equivalent to the corresponding quadratic form under the vectorized version, but it avoids constructing the full $pq\times pq$ precision matrix.

Define the gradient matrix
\begin{equation*}
    \mathbf{G} = \boldsymbol{\Xi}_1\left(\hat{\mathbf{Y}}-{\mathbf{M}}\right)\boldsymbol{\Xi}_2. 
\end{equation*}
Consider a missing entry $\hat{Y}_{ij}$ in row $i$ and column $j$, while holding all other entries of $\hat{\mathbf{Y}}$ fixed. The first derivative of $J(\hat{\mathbf{Y}})$ with respect to $\hat{Y}_{ij}$ is $\mathbf{G}_{ij}$, whereas the corresponding second derivative is $(\boldsymbol{\Xi}_1)_{ii}(\boldsymbol{\Xi}_2)_{jj}$. Therefore, the exact minimizer of the objective with respect to this individual entry is obtained using the displacement
\begin{equation*}
    \eta_{ij} = -\frac{\mathbf{G}_{ij}}{(\boldsymbol{\Xi}_1)_{ii}(\boldsymbol{\Xi}_2)_{jj}},
\end{equation*}
followed by the update
\begin{equation} \label{eq:meanupdate}
    \hat{Y}_{ij} = \hat{Y}_{ij} + \eta_{ij}.
\end{equation}

Applying \eqref{eq:meanupdate} sequentially to all missing entries constitutes a single Gauss-Seidel sweep. Each coordinate is updated using the most recent values of the other missing entries, ensuring that every update minimizes the objective with respect to the active coordinate and does not increase $J(\hat{\mathbf{Y}})$. When model parameters and observed entries are held fixed, repeated sweeps converge to the unique minimizer of $J(\hat{\mathbf{Y}})$, which corresponds to the exact conditional mean of the missing entries. In the matrix-normal setting, the row and column covariance factors are estimated through flip-flop updates, where each factor is updated conditionally on the current value of the other. As a result, the conditional-mean objective changes after each covariance-factor update, making it generally unnecessary to solve the current conditional-mean problem to full convergence at every iteration. Accordingly, a single Gauss-Seidel sweep is performed, initialized using the imputed values from the preceding iteration. This approach yields a computationally efficient partial E-step that decreases $J(\hat{\mathbf{Y}})$, while subsequent flip-flop iterations progressively refine both the imputed values and the covariance estimates.

\subsection{E-step for conditional covariance} \label{sec:covupdate}
We next consider the part of the KL-divergence objective that depends on the conditional covariance of the missing entries. Let $m$ represent the number of missing entries, $\boldsymbol{\Xi}_{zz}$ denote the $m \times m$ submatrix of the full precision matrix corresponding to the missing coordinates, and $\hat{\mathbf{Z}}$ represent the current estimate of their conditional covariance matrix. Up to an additive constant, considering \eqref{eq:Schur}, the covariance-dependent objective is
\begin{equation} \label{eq:covloss}
    J(\hat{\mathbf{Z}}) = \operatorname{tr}\{\boldsymbol{\Xi}_{zz}\hat{\mathbf{Z}}\}-\ln |\hat{\mathbf{Z}}|.
\end{equation}

To derive a coordinate update, consider one missing coordinate indexed by $w$. Without loss of generality, reorder the missing coordinates so that the active coordinate appears first, and partition the precision and covariance matrices as
\begin{equation*}
    \boldsymbol{\Xi}_{zz} = \begin{bmatrix} \xi_{w} & \boldsymbol{\xi}_w^\prime \\ \boldsymbol{\xi}_w & \boldsymbol{\Xi}_{-w,-w} \end{bmatrix}, \quad \hat{\mathbf{Z}} = \begin{bmatrix} w & \mathbf{w}^\prime \\ \mathbf{w} & \mathbf{W} \end{bmatrix},
\end{equation*}
where $\xi_{w}$ is the diagonal precision associated with coordinate $w$, $\boldsymbol{\xi}_w$ contains its cross-precision terms, $w$ is its current conditional variance, $\mathbf{w}$ contains its cross-covariances with the remaining missing coordinates, and $\mathbf{W}$ is the covariance matrix of the remaining coordinates. In the active coordinate update, $\mathbf{W}$ is held fixed while $w$ and $\mathbf{w}$ are optimized jointly.  
Using the block structure above, the trace term in \eqref{eq:covloss} becomes
\begin{equation*}
    \operatorname{tr}(\boldsymbol{\Xi}_{zz} \hat{\mathbf{Z}}) = \xi_{w} w + 2\boldsymbol{\xi}_w^\prime \mathbf{w} + \operatorname{tr}(\boldsymbol{\Xi}_{-w,-w} \mathbf{W}).
\end{equation*}
Because the final term does not depend on the active variables, it can be regarded as a constant. Moreover, the determinant gives
\begin{equation*}
    |\hat{\mathbf{Z}}| = |\mathbf{W}| \left( w - \mathbf{w}^\prime \mathbf{W}^{-1} \mathbf{w} \right).
\end{equation*}
Define the Schur complement scalar as $s = w - \mathbf{w}^\prime \mathbf{W}^{-1} \mathbf{w}$. After removing terms that are constant with respect to $w$ and $\mathbf{w}$, the active objective reduces to
\begin{equation*}
    J(w, \mathbf{w}) = \frac{1}{2} \left( \xi_w w + 2\boldsymbol{\xi}_w^\prime \mathbf{w} - \log s \right),
\end{equation*}
subject to $s > 0$. To find the optimal target variance, we take the partial derivative of $J$ with respect to $w$, which is given by
\begin{equation*}
    \frac{\partial J}{\partial w} = \frac{1}{2} \left( \xi_w - \frac{1}{s} \right).
\end{equation*}
Thus, the first-order condition for the target variance is $s = 1/\xi_w.$ We now optimize with respect to the cross-covariance vector $\mathbf{w}$. Using the chain rule, the gradient of the Schur complement is $\nabla_{\mathbf{w}} s = -2\mathbf{W}^{-1} \mathbf{w}$. Thus, the gradient with respect to the cross-covariance vector is
\begin{equation*}
    \nabla_{\mathbf{w}} J = \frac{1}{2} \left[ 2\boldsymbol{\xi}_w - \frac{1}{s} (-2\mathbf{W}^{-1} \mathbf{w}) \right] = \boldsymbol{\xi}_w + \frac{1}{s} \mathbf{W}^{-1} \mathbf{w}.
\end{equation*}
Combining this with $s = 1/\xi_w$ yields the cross-covariance $\mathbf{w}$ update
\begin{equation}\label{eq:crosscovupdate}
    \hat{\mathbf{w}} = -s \mathbf{W} \boldsymbol{\xi}_w = -\frac{1}{\xi_w} \mathbf{W} \boldsymbol{\xi}_w.
\end{equation}
Finally, substituting \eqref{eq:crosscovupdate} into the Schur-complement condition gives the corresponding variance update
\begin{equation}\label{eq:covupdate}
    \hat{w} = \frac{1}{\xi_w} + \hat{\mathbf{w}}^\prime \mathbf{W}^{-1} \hat{\mathbf{w}} = \frac{1}{\xi_w} + \frac{1}{\xi_w^2} \boldsymbol{\xi}_w^\prime \mathbf{W} \boldsymbol{\xi}_w.
\end{equation}
Although $\mathbf{W}^{-1}$ appears in the derivation, the implementations \eqref{eq:covupdate} and \eqref{eq:crosscovupdate} require only matrix products and do not require explicitly inverting $\mathbf{W}$.

One covariance sweep is obtained by applying these updates sequentially to every missing coordinate. If $\mathbf{W} \succ 0$, then the updated Schur complement satisfies $\hat{s} = \xi_w^{-1} > 0$, because $\boldsymbol{\Xi}_{zz} \succ 0$ implies $\xi_w > 0$. Consequently, each update preserves the positive definiteness of $\hat{\mathbf{Z}}$. Moreover, the updated row and column exactly minimize the active objective while the remaining covariance block is held fixed. Each coordinate update therefore cannot increase $J(\hat{\mathbf{Z}})$ and decreases it unless the active block already satisfies its first-order optimality conditions. At a fixed point of the coordinate updates, the first-order conditions hold for every row and column, implying $\hat{\mathbf{Z}}^{-1} = \mathbf{\Xi}_{zz}$. Because the objective in \eqref{eq:covloss} is strictly convex over the cone of positive-definite matrices, this minimizer is unique. Accordingly, one sweep constitutes a partial covariance update, whereas repeated sweeps recover the exact conditional covariance.

The proposed procedure operates only on the coordinates associated with the missing entries. In particular, for missing cells $a=(i_a,i_b)$ and $b=(i_a,i_b)$ the corresponding entry of the conditional precision matrix is obtained directly from the Kronecker factors as $(\boldsymbol{\Xi}_{zz})_{ab} = (\boldsymbol{\Xi}_1)_{(i_a,i_b)}(\boldsymbol{\Xi}_2)_{(i_a,i_b)}$. Thus, neither the full $pq\times pq$ precision matrix nor the full matrix-normal covariance matrix needs to be constructed. In addition, unlike a direct exact E-step based on solving or inverting the complete missing-coordinate system, each block update uses only the current $m\times m$ covariance estimate and the corresponding entries of $\boldsymbol{\Xi}_{zz}$. This retains the fixed point of the exact E-step while allowing the conditional moments to be updated incrementally through partial Gauss-Seidel sweeps.

\subsection{M-step}
Given the conditional expectation $\hat{\mathbf{Y}}_i$ and covariance $\hat{\mathbf{V}}_i$, we can update the distribution parameters $\mathbf{M}$, $\boldsymbol{\Sigma}_1$, and $\boldsymbol{\Sigma}_2$.First, the location matrix is updated as the empirical mean
\begin{equation} \label{eq:parammean}
    \hat{\mathbf{M}} = \frac{1}{N}\sum_{i=1}^N\hat{\mathbf{Y}}_i.
\end{equation}
To update the covariance matrices, following the discussion in Section \ref{sec:emformat}, we project $\hat{\mathbf{V}}_i$ onto the row and column spaces. For the row covariance, we compute a correction matrix $\hat{\mathbf{R}}_i$ to capture the missing variance. Its $(k,l)$-th entry is obtained by weighting the corresponding row-space sub-block of the conditional covariance, denoted as $\hat{\mathbf{V}}_{i, kl}^\text{R}$, by the column precision matrix $\boldsymbol{\Xi}_2= \boldsymbol{\Sigma}_2^{-1}$, which can be expressed as
\begin{equation*}
    \hat{\mathbf{R}}_{i, kl} = \operatorname{tr}\left(\boldsymbol{\Xi}_2 \hat{\mathbf{V}}_{i, kl}^\text{R}\right).
\end{equation*}
Combining this correction term with the empirical outer product of the expected sufficient statistics yields the unconstrained row covariance $\widetilde{\boldsymbol{\Sigma}}_1$. This is then normalized to resolve scale identifiability
\begin{equation} \label{eq:paramrow}
    \widetilde{\boldsymbol{\Sigma}}_1 = \frac{1}{N q} \sum_{i=1}^N \left[ (\hat{\mathbf{Y}}_i - \hat{\mathbf{M}}) \boldsymbol{\Xi}_2 (\hat{\mathbf{Y}}_i - \hat{\mathbf{M}})^\prime + \hat{\mathbf{R}}_i \right], \quad \hat{\boldsymbol{\Sigma}}_1 = \frac{\widetilde{\boldsymbol{\Sigma}}_1}{|\widetilde{\boldsymbol{\Sigma}}_1|^{1/p}},
\end{equation}
where $p$ and $q$ are the number of rows and columns, respectively. By symmetry, we define a column covariance correction matrix $\hat{\mathbf{C}}_i$, projecting onto the column space using the row precision matrix $\boldsymbol{\Xi}_1$ via
\begin{equation*}
    \hat{\mathbf{C}}_{i, kl} = \operatorname{tr}\left(\boldsymbol{\Xi}_1 \hat{\mathbf{V}}_{i, kl}^\text{C}\right),
\end{equation*}
where $\hat{\mathbf{V}}_{i, kl}^\text{C}$ represents the $(k,l)$-th column-space sub-block of $\hat{\mathbf{V}}_i$. The column covariance matrix is updated and normalized similarly
\begin{equation} \label{eq:paramcol}
    \widetilde{\boldsymbol{\Sigma}}_2 = \frac{1}{N p} \sum_{i=1}^N \left[ (\hat{\mathbf{Y}}_i - \hat{\mathbf{M}})^\prime \boldsymbol{\Xi}_1 (\hat{\mathbf{Y}}_i - \hat{\mathbf{M}}) + \hat{\mathbf{C}}_i \right], \quad \hat{\boldsymbol{\Sigma}}_2 = \frac{\widetilde{\boldsymbol{\Sigma}}_2}{|\widetilde{\boldsymbol{\Sigma}}_2|^{1/q}}.
\end{equation}
This sequential formulation cleanly separates the imputation of the missing variance via the correction terms from the standard structural covariance updates. Finally, since $\widetilde{\boldsymbol{\Sigma}}_2$ is updated using the newly updated $\hat{\boldsymbol{\Sigma}}_1$, the scalar variance parameter is updated as
\begin{equation*}
    \hat{\sigma}^2 = |\widetilde{\boldsymbol{\Sigma}}_2|^{1/q}.
\end{equation*}

\subsection{Structural missingness} \label{sec:structural}
Previously, only the arbitrary element-wise missingness was considered. In fact, with different structural missing patterns, the separability will remain to a different extent. Here, we consider the case in which the missing entries form a submatrix in each matrix. Following the notation from previous sections, with such a missing pattern, the submatrix of the precision matrix can be written as
\begin{equation*}
    \boldsymbol{\Xi}_{zz} = \sigma^{-2}\mathbf{K}_2 \otimes \mathbf{K}_1,
\end{equation*}
where $\mathbf{K}_1$ and $\mathbf{K}_2$ are the corresponding principal submatrices of the current precision matrices. Consequently, the exact conditional covariance of the missing block is also separable. We therefore maintain the representation 
\begin{equation*}
    \hat{\mathbf{Z}} = \sigma^2 \hat{\mathbf{Z}}_2 \otimes \hat{\mathbf{Z}}_1
\end{equation*}
throughout the partial updates. With such a structure, $\hat{\mathbf{Z}}_1$ and $\hat{\mathbf{Z}}_2$ can be updated by independently performing \eqref{eq:crosscovupdate} and \eqref{eq:covupdate}. Let $\mathcal{R} = \{r_1, \dots, r_{p^*}\} \subseteq \{1, \dots, p\}$ and $\mathcal{C} = \{c_1, \dots, c_{q^*}\} \subseteq \{1, \dots, q\}$ denote the row and column indices of the rectangular missing block in observation $i$, respectively. Define the corresponding embedding matrices $\mathbf{E}_\mathcal{R} = [\mathbf{e}_{r_1}, \dots, \mathbf{e}_{r_{p^*}}] \in \mathbb{R}^{p \times p_i}$ and $\mathbf{E}_\mathcal{C} = [\mathbf{e}_{c_1}, \dots, \mathbf{e}_{c_{q^*}}] \in \mathbb{R}^{q \times q_i}$, where $\mathbf{e}_j$ denotes the appropriate standard basis vector. The row- and column-space covariance factors are embedded into the full matrix dimensions as
\begin{equation*}
    \hat{\mathbf{V}}^R = \mathbf{E}_\mathcal{R} \hat{\mathbf{Z}}_{1i} \mathbf{E}_\mathcal{R}' \in \mathbb{R}^{p \times p}, \quad \hat{\mathbf{V}}^C = \mathbf{E}_\mathcal{C} \hat{\mathbf{Z}}_{2i} \mathbf{E}_\mathcal{C}' \in \mathbb{R}^{q \times q}.
\end{equation*}
Thus, $(\hat{\mathbf{V}}^R)_{r_{a}, r_{b}} = (\hat{\mathbf{Z}}_{1})_{ab}$ and $(\hat{\mathbf{V}}^C)_{c_{a}, c_{b}} = (\hat{\mathbf{Z}}_{2})_{ab}$, while all entries outside the corresponding missing-row and missing-column index sets are zero. Under the column-wise vectorization convention, the resulting covariance correction in the full $pq$-dimensional space is
\begin{equation*}
    \hat{\mathbf{V}} = \sigma^2 \hat{\mathbf{V}}^C \otimes \hat{\mathbf{V}}^R.
\end{equation*}
Consequently, the projected correction terms can be written as
\begin{equation*}
    \hat{\mathbf{R}} = \sigma^2 \text{tr}\left(\mathbf{K}_{2} \hat{\mathbf{Z}}_{2}\right) \hat{\mathbf{V}}^R, \quad \hat{\mathbf{C}} = \sigma^2 \text{tr}\left(\mathbf{K}_{1} \hat{\mathbf{Z}}_{1}\right) \hat{\mathbf{V}}^C,
\end{equation*}
where $\mathbf{K}_{1} = \mathbf{\Xi}_1[\mathcal{R}, \mathcal{R}]$ and $\mathbf{K}_{2} = \mathbf{\Xi}_2[\mathcal{C}, \mathcal{C}]$.


\subsection{Simulation studies for efficiency comparison} \label{sec:sim1}
This simulation compares the proposed method with three alternative matrix normal missing-data estimators under both the missing completely at random (MCAR) condition and the structural missingness described in Section~\ref{sec:structural}. The first alternative is the standard exact EM algorithm for matrix-variate normal data \citep{glanz2018expectation}, which leverages the exact conditional distribution of missing entries in the E-step and serves as our benchmark for runtime, likelihood, and estimation accuracy. The second is a plug-in mean-imputation method that replaces missing entries with their conditional-mean approximations but omits conditional covariance corrections. Finally, the third method is a hybrid approach that retains the exact E-step of the standard EM algorithm but applies the M-step proposed by \cite{lachos2025algorithm}. For the submatrix missingness pattern only, we additionally include the block version of our method, which exploits the separability of the missing block and updates the two Kronecker factors of the conditional covariance independently. 

To ensure a fair comparison, we directly use the implementation from \cite{glanz2018expectation} for both the exact EM and plug-in methods. For the third alternative, we retain the exact EM's E-step but utilize a literal implementation for the M-step. Our proposed method follows a similar structural scheme to the exact EM. It is implemented in \texttt{R}, with computationally intensive components written in \texttt{C}. All simulations were performed on a machine equipped with an Apple M4 chip and 16 GB of memory.

For each replicate, the complete data are generated from a matrix normal model
\begin{equation*}
    \mathcal{Y} \sim \mathcal{N}_{p\times q}(\mathbf{M},\boldsymbol{\Sigma}_1,\boldsymbol{\Sigma}_2).
\end{equation*}
The mean matrix is fixed across replicates and has entries determined by a smooth sinusoidal pattern, which is
\begin{equation*}
    \mathbf{M} = \frac{1}{3}\operatorname{matrix}\left[\sin\left(\frac{1}{5}\right),\sin\left(\frac{2}{5}\right),\ldots,\sin\left(\frac{pq}{5}\right)\right],
\end{equation*}
where $\operatorname{matrix}$ is the operator which folds the vector back into a $p \times q$ matrix. The row covariance matrix $\boldsymbol{\Sigma}_1$ follows an $\operatorname{AR}(1)$ correlation structure with correlation $\rho_1 = 0.55$, and the column covariance matrix $\boldsymbol{\Sigma}_2$ follows an $\operatorname{AR}(1)$ with correlation parameter $\rho_2 = 0.45$. For two entries $a=(i_a,j_a)$ and $b=(i_b,j_b)$, the covariance are calculated via
\begin{equation*}
    (\boldsymbol{\Sigma}_1)_{i_a,i_b} = \rho_1^{|i_a-i_b|}, \quad \text{and }(\boldsymbol{\Sigma}_2)_{j_a,j_b} = \rho_2^{|j_a-j_b|}.
\end{equation*}
And the scale parameter is $\sigma^2 = 1.2$.

The sample size is fixed at $N=1000$ for all dimensional settings. This choice is consistent with the simulation design of \citet{glanz2018expectation}, who considered sample sizes ranging from 500 to 2000 for matrix dimensions up to $10\times25$. Our simulation settings are $(p,q) \in \{(6,9),(12,16),(15,20),(21,24)\}$, with the largest setting containing 504 variables after vectorization, compared with 250 in their largest setting. Because the primary objective of this study is to evaluate how computational efficiency changes with matrix dimensionality, $N$ is held fixed so that the observed differences are not confounded by changes in sample size. For the random missing setting, entries are missing independently, with target missing proportions of $10\%, 25\%, 50\%, 75\%$. 

For the structural missingness experiments, the $6\times 9$ setting is omitted because the matrix is too small for the rectangular submatrix structure to meaningfully demonstrate the computational advantage of the block-based approximation. Thus, the structural missingness comparison is conducted for $(p,q)\in\{(12,16),(15,20),(21,24)\}$. For each target missing proportion, the missing submatrix size is selected by minimizing the discrepancy between the block area and the target number of missing entries. Specifically, the block sizes are $4\times5$, $6\times8$, $8\times12$, and $12\times12$ for the $12\times16$ setting; $5\times6$, $5\times15$, $10\times15$, and $15\times15$ for the $15\times20$ setting; and $5\times10$, $9\times14$, $14\times18$, and $18\times21$ for the $21\times24$ setting, corresponding respectively to the target missing proportions $10\%,25\%,50\%$, and $75\%$. For both settings, 30 repeats are conducted.

Following \cite{glanz2018expectation}, convergence is assessed by the relative change in parameters. At iteration $t$, we stop when
$$\frac{\|\hat{\mathbf{M}}^{(t+1)} - \hat{\mathbf{M}}^{(t)}\|_1}{\|\hat{\mathbf{M}}^{(t)}\|_1} + \frac{\|\hat{\boldsymbol{\Sigma}}_2^{(t+1)} - \hat{\boldsymbol{\Sigma}}_2^{(t)}\|_{1}}{\|\hat{\boldsymbol{\Sigma}}_2^{(t)}\|_{1}} + \frac{\|\hat{\boldsymbol{\Sigma}}_1^{(t+1)} - \hat{\boldsymbol{\Sigma}}_1^{(t)}\|_{1}}{\|\hat{\boldsymbol{\Sigma}}_1^{(t)}\|_{1}} + \frac{|\hat{\sigma}^{2(t+1)} - \hat{\sigma}^{2(t)}|}{|\hat{\sigma}^{2(t)}|}\leq\epsilon,$$
where $\|\cdot\|_1$ means the sum of absolute values. In the simulations, $\epsilon=10^{-5}$.
\begin{figure}
    \centering
    \includegraphics[width=1\linewidth]{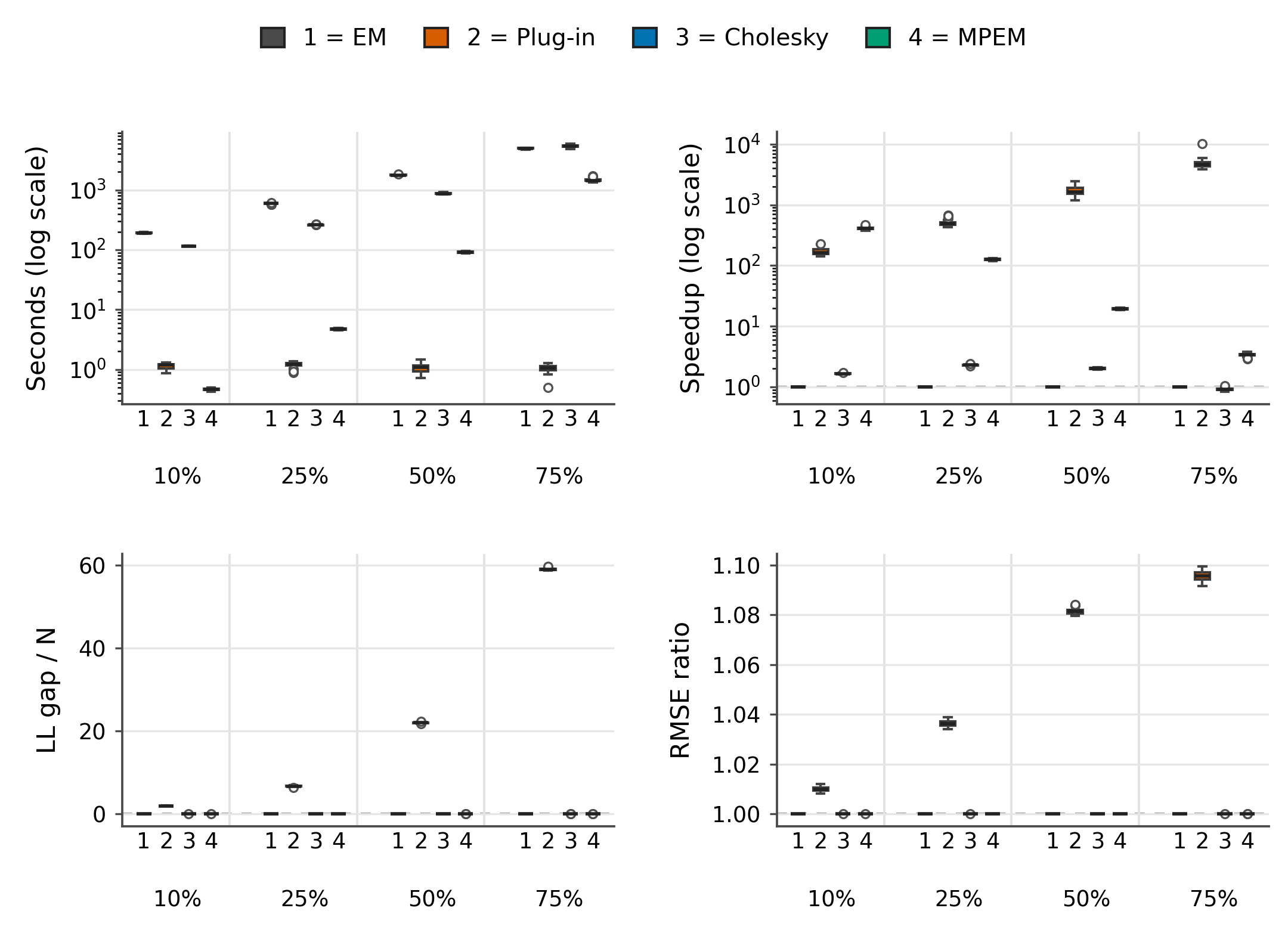}
    \caption{Performance comparison of four computational methods (EM, Plug-in, Cholesky, and MPEM) across varying proportions of missing data (10\%, 25\%, 50\%, and 75\%) with the dimensionality $15\times 20$. The top row evaluates computational efficiency, displaying the absolute runtime in seconds (top left) and the relative speedup (top right), both shown on a logarithmic scale. The bottom row assesses statistical accuracy, depicting the log-likelihood gap normalized by sample size (LL gap / N, bottom left) and the Root Mean Square Error (RMSE) ratio (bottom right).}
    \label{fig:1}
\end{figure}
\begin{figure}
    \centering
    \includegraphics[width=1\linewidth]{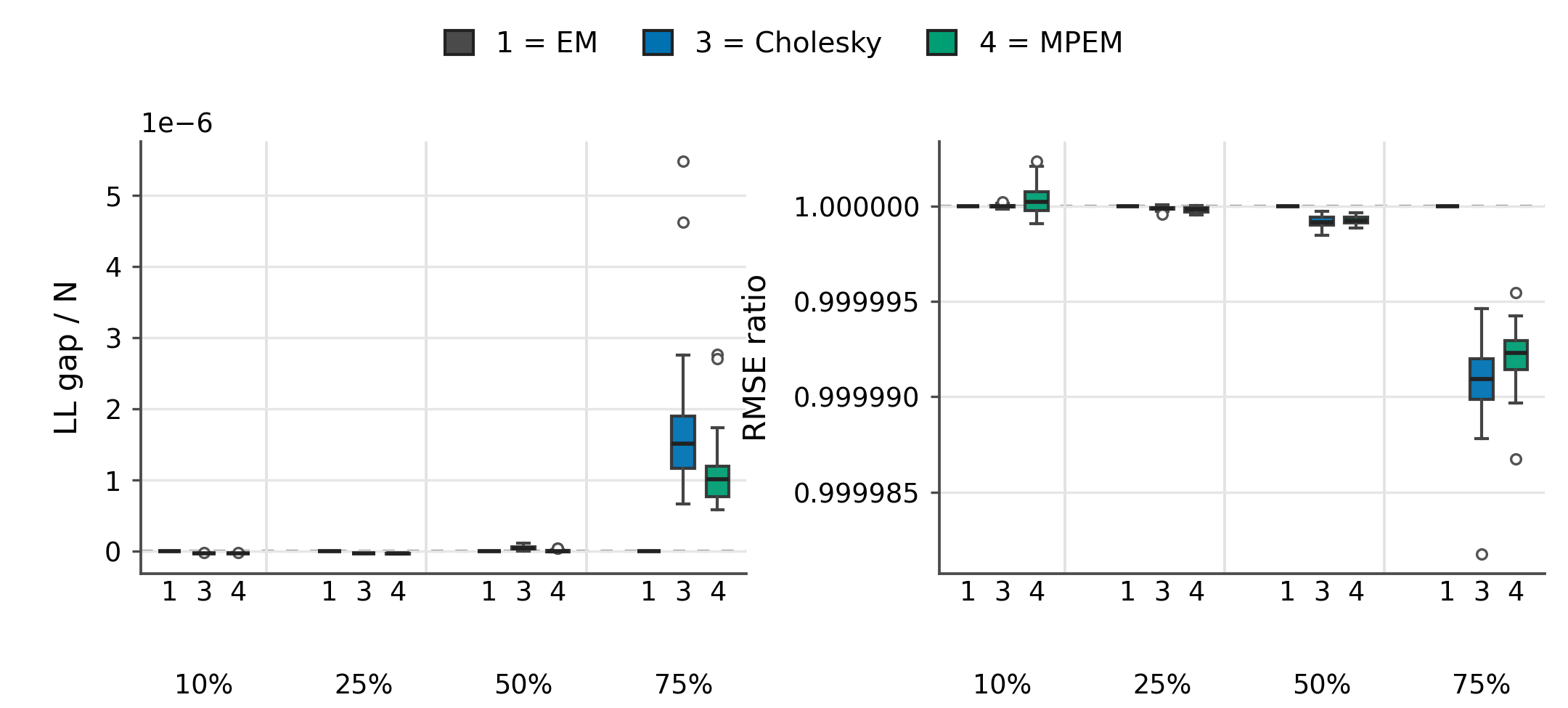}
    \caption{A magnified view of the statistical accuracy metrics comparing the EM, Cholesky, and MPEM across varying missing data proportions (10\% to 75\%). The left panel details the normalized log-likelihood gap (LL gap / N) on a micro-scale ($10^{-6}$), while the right panel displays the RMSE ratio with a highly restricted y-axis.}
    \label{fig:2}
\end{figure}
\begin{figure}
    \centering
    \includegraphics[width=1\linewidth]{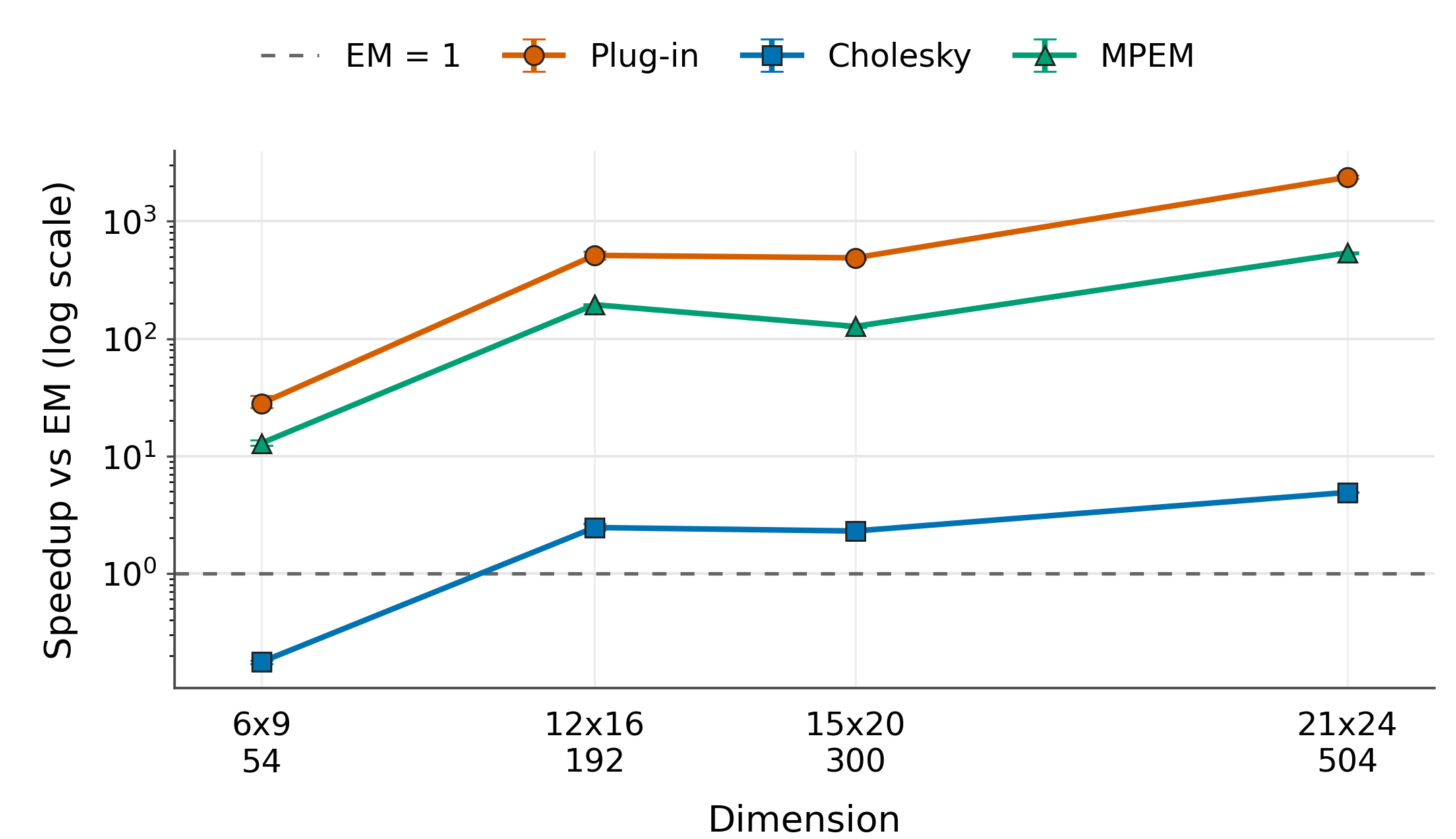}
    \caption{Relative computational speedup of the Plug-in, Cholesky, and MPEM compared to the Exact baseline, averaged over all simulation replicates. The y-axis displays the speedup multiplier on a logarithmic scale, with the Exact method serving as the reference at $1$ (dashed line). The x-axis tracks increasing problem dimensions, ranging from $6 \times 9$ (54 total elements) to $21 \times 24$ (504 total elements).}
    \label{fig:3}
\end{figure}
\begin{figure}
    \centering
    \includegraphics[width=1\linewidth]{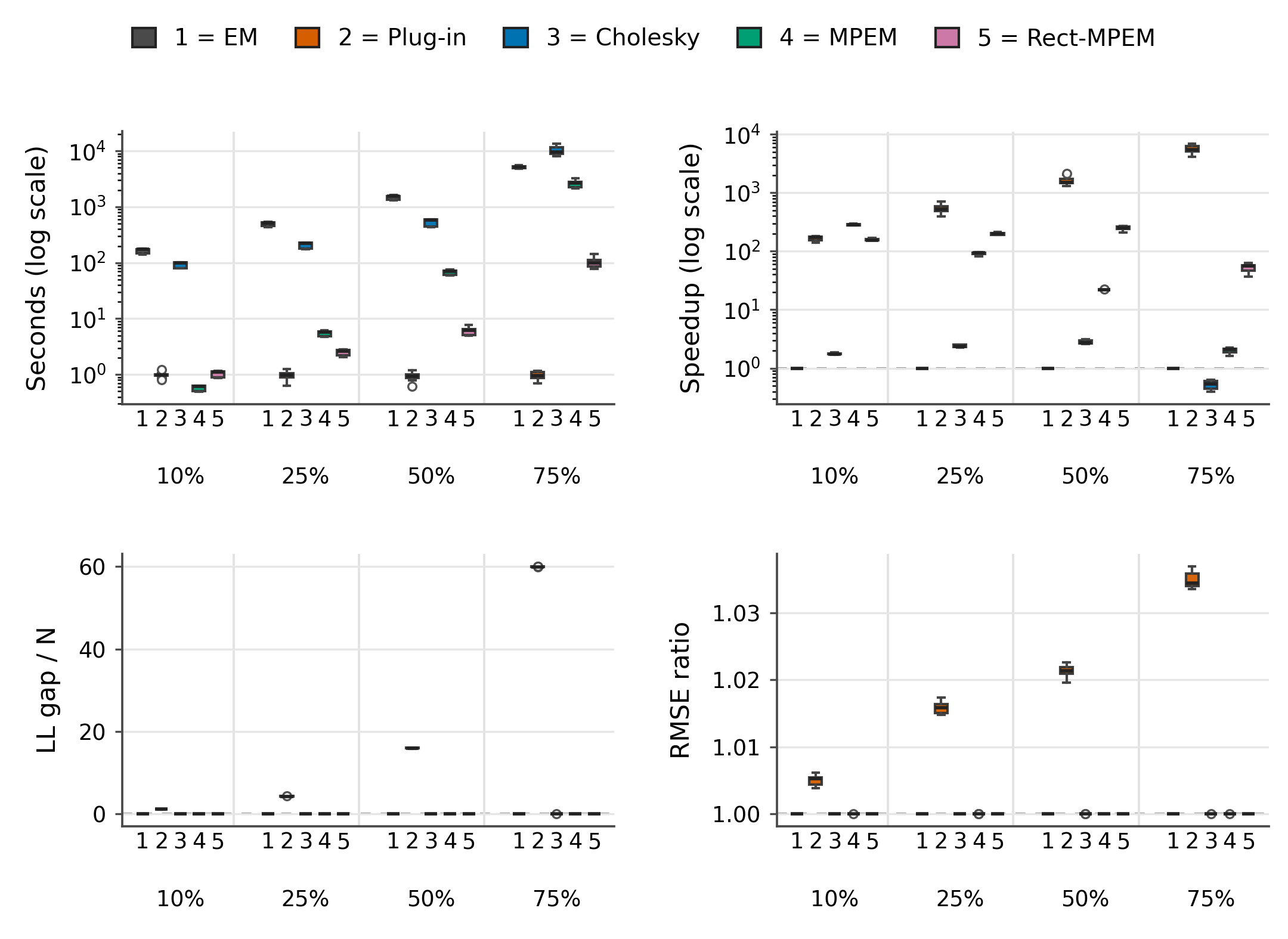}
    \caption{Performance comparison of five computational methods (EM, Plug-in, Cholesky, MPEM, and Rect-MPEM) with the missing proportion 25\%, dimensionality $15\times 20$, and the structural missing pattern.}
    \label{fig:4}
\end{figure}
In Figure~\ref{fig:1}, Figure~\ref{fig:2}, and Figure~\ref{fig:3}, the log-likelihood (LL) gap is computed as the observed log-likelihood of EM minus that of each competing method, divided by the number of observations \(N\). The root mean square error (RMSE) ratio is computed as each method’s missing-entry imputation RMSE divided by the corresponding RMSE of EM, so values near one indicate EM-level imputation accuracy. With arbitrary missingness, the simulation results indicate that the proposed method substantially reduces computation time compared with the standard exact EM algorithm, while maintaining nearly the same observed-data likelihood across most settings. This is expected because exact EM repeatedly computes the full conditional distribution of the missing entries, whereas our method avoids the expensive full conditional covariance calculation and instead updates the missing conditional moments approximately. The plug-in method is usually the fastest, but it ignores the missing-data covariance correction and can therefore lead to a noticeable loss in likelihood or parameter accuracy. For the M-step with the Cholesky factorization, since the Cholesky factors are computed from the full outer product and conditional covariance, when dimensionality is low, it is slower than the exact EM. But as dimensionality increases, Cholesky factorization offers some efficiency gains, though much slower than MPEM. Thus, the main comparison is between exact EM, the likelihood benchmark, and our method, a faster approximation with much lower computational cost. For the submatrix missing pattern, we conduct the comparison under the case of $15\times 20$ and a $25\%$ missing rate. In this case, our structural imputation approach gains a larger improvement in efficiency.

\section{Missingness in mixtures of matrix normals}\label{sec:mixtures}
The EM algorithm also provides a standard approach for fitting finite mixture models, in which component membership is latent. In the context of finite mixture models, the EM algorithm can jointly perform clustering and component-specific imputation.

\subsection{Mixtures of matrix-variate normal distributions}
A natural extension of the matrix normal distribution is the mixture of matrix normal distributions (MMN) introduced by \cite{viroli2011finite}. With the constraint discussed in Section~\ref{sec:identifiability}, the probability density function of the matrix normal distribution can be written as
\begin{equation*}
    \phi_{p \times q}(\mathbf{Y} \mid \mathbf{M},\boldsymbol{\Sigma}_1,\boldsymbol{\Sigma}_2, \sigma^2) = \frac{\operatorname{exp}\left\{-\frac{1}{2\sigma^2}\operatorname{tr}\left(\boldsymbol{\Sigma}_1^{-1}(\mathbf{Y}-\mathbf{M})\boldsymbol{\Sigma}_2^{-1}(\mathbf{Y}-\mathbf{M})^\prime\right)\right\}}{(2\pi\sigma^2)^{\frac{pq}{2}}|\boldsymbol{\Sigma}_1|^{\frac{q}{2}}|\boldsymbol{\Sigma}_2|^{\frac{p}{2}}}. 
\end{equation*}
Using the matrix normal distribution as the component density, the density of a $G$-component MMN is
\begin{equation} \label{eq: MMN density}
    f(\mathbf{Y}\mid \boldsymbol{\vartheta}) = \sum_{g=1}^G \pi_g\phi_{ p \times q}(\mathbf{Y} \mid \mathbf{M}_g,\boldsymbol{\Sigma}_{1g},\boldsymbol{\Sigma}_{2g}, \sigma^2_g),
\end{equation}
where $\phi_{ p \times q}(\mathbf{Y} \mid \mathbf{M}_g,\boldsymbol{\Sigma}_1,\boldsymbol{\Sigma}_2, \sigma^2_g)$ denotes the $g$th matrix-variate normal density function, and $\boldsymbol{\vartheta}$ is the parameter space. Following this mixture of matrix normal distributions, a substantial number of matrix-variate mixture models have been proposed in recent years \citep{dougru2016finite,gallaugher2018finite,tomarchio2022mixtures, sharp2023dual, silva2023finite,tomarchio2026matrix}.

For parameter estimation, the main difference from the single-component case is that in the E-step, not only are the central moments of the missing component estimated, but also the latent membership. Let $u_{ig}$ denote the latent membership so that $u_{ig}=1$ if $\mathbf{Y}_i$ belongs to group $g$. The conditional expectation of $u_{ig}$ is computed from the exact marginal density of the observed entries rather than from the density evaluated at an imputed complete matrix. Let $\mathbf{\Sigma}_g = \sigma_g^2 (\mathbf{\Sigma}_{2g} \otimes \mathbf{\Sigma}_{1g})$ and $\boldsymbol{\Xi}_g = \mathbf{\Sigma}_g^{-1}$, and $\boldsymbol{\Sigma}^i_{g,x}$ denote the observed covariance block of the observation $i$, and $\boldsymbol{\Xi}^i_{g,z}$ denote the corresponding missing precision block. Using the determinant identity
\begin{equation*}
    \log |\boldsymbol{\Sigma}^i_{g,x}| = \log |\mathbf{\Sigma}_g| + \log |\boldsymbol{\Xi}^i_{g,z}|,
\end{equation*}
the observed-data density can be evaluated through the missing precision block. Specifically, let $\hat{\mathbf{Y}}_{ig}$ denote the completed matrix, then the observed quadratic form is obtained exactly as
\begin{equation*}
    Q_{ig} = \frac{1}{\sigma^2_g}\operatorname{tr}\left(\boldsymbol{\Xi}_{1g}(\hat{\mathbf{Y}}_{ig}-\mathbf{M}_g)\boldsymbol{\Xi}_{2g}(\hat{\mathbf{Y}}_{ig}-\mathbf{M}_g)^\prime\right),
\end{equation*}
and hence
\begin{equation*}
    \log \phi(\mathbf{y}_{ig,x}\mid \boldsymbol{\vartheta}_g) = -\frac{1}{2} \left[ o \log(2\pi) + \log |\mathbf{\Sigma}_g| + \log |\boldsymbol{\Xi}_{ig,z}| + Q_{ig} \right],
\end{equation*}
where $\mathbf{y}_{ig,x}$ is the vectorized observed component of $\hat{\mathbf{Y}}_{ig}$, $o$ is the number of observed entries, and $\boldsymbol{\vartheta}_g$ is the parameter space of $g$th component. Note that $|\mathbf{\Sigma}_g| = pq\log \sigma^2_g$ with the determinant constraints. For the submatrix missingness, the Kronecker factorization remains, so the evaluation of $\phi(\mathbf{y}_{ig,x}\mid \boldsymbol{\vartheta}_g)$ becomes simpler. Then the membership probabilities are therefore updated by
\begin{equation*}
    \hat{u}_{i g}=\frac{\pi_g\phi(\mathbf{y}_{ig,x}\mid \boldsymbol{\vartheta}_g)}{\sum_{h=1}^G \pi_h \phi(\mathbf{y}_{ih,x}\mid \boldsymbol{\vartheta}_h)}.
\end{equation*}
Within each component, for each observation $\mathbf{Y}_i$, we perform \eqref{eq:meanupdate} and \eqref{eq:covupdate} to obtain the component-specific sufficient statistics $\hat{\mathbf{Y}}_{ig}$ and $\hat{\mathbf{V}}_{ig}$. The update for the mixing proportion is $\hat{\pi}_g = N_g/N$, where $N_g = \sum_{i=1}^N \hat{u}_{ig}$. Then, the distribution parameters are updated via the following weighted versions of \eqref{eq:parammean}, \eqref{eq:paramrow}, and \eqref{eq:paramcol}. For the component mean,
\begin{equation*}
    \hat{\mathbf{M}}_g = \frac{1}{N_g}\sum_{i=1}^N \hat{u}_{ig}\hat{\mathbf{Y}}_{ig}.
\end{equation*}
For the covariance matrices, 
\begin{equation}\label{eq:rowcov}
    \widetilde{\boldsymbol{\Sigma}}_{1g} = \frac{1}{N_g q} \sum_{i=1}^N \hat{u}_{ig} \left[ (\hat{\mathbf{Y}}_{ig} - \hat{\mathbf{M}}_g) \boldsymbol{\Xi}_{2g} (\hat{\mathbf{Y}}_{ig} - \hat{\mathbf{M}}_g)^\prime + \hat{\mathbf{R}}_{ig} \right], \quad \hat{\boldsymbol{\Sigma}}_{1g} = \frac{\widetilde{\boldsymbol{\Sigma}}_{1g}}{|\widetilde{\boldsymbol{\Sigma}}_{1g}|^{1/p}},
\end{equation}
where $\hat{\mathbf{R}}_{ig}$ is the component-specific correction term projected on the row space, and
\begin{equation} \label{eq:colcov}
    \widetilde{\boldsymbol{\Sigma}}_{2g} = \frac{1}{N_g p} \sum_{i=1}^N \hat{u}_{ig}\left[ (\hat{\mathbf{Y}}_{ig} - \hat{\mathbf{M}}_g)^\prime \boldsymbol{\Xi}_{1g} (\hat{\mathbf{Y}}_{ig} - \hat{\mathbf{M}}_g) + \hat{\mathbf{C}}_{ig} \right], \quad \hat{\boldsymbol{\Sigma}}_{2g} = \frac{\widetilde{\boldsymbol{\Sigma}}_{2g}}{|\widetilde{\boldsymbol{\Sigma}}_{2g}|^{1/q}},
\end{equation}
where $\hat{\mathbf{C}}_{ig}$ is the component-specific correction term projected on the column space. For the variance scalar, 
\begin{equation} \label{eq:scalarvar}
    \hat{\sigma}^2_{g} = |\widetilde{\boldsymbol{\Sigma}}_{2g}|^{1/q}.
\end{equation}
Once the algorithm reaches convergence, the $i$th full imputed matrix is defined as 
\begin{equation*}
    \hat{\mathbf{Y}}_{i} = \sum_{g=1}^G \hat{u}_{ig}\hat{\mathbf{Y}}_{ig}.
\end{equation*}

\subsection{Mixtures of spatial factor analyzers}
In finite normal mixture models, model complexity is often dominated by the covariance parameterization, as the number of free covariance parameters grows quadratically with the dimensionality. Moreover, for certain types of data, specific constraints can be imposed on the covariance matrices. In the context of spatial data, building on the linear spatial correlation model introduced by \cite{worsley1991linear}, \cite{lu2026spatialcovarianceconstraintsgaussian} developed a sigmoid decay (SD) spatial covariance structure that assumes spatial covariance decreases with distance according to a parametric sigmoid function. This assumption, however, is often too restrictive in practical applications, and estimation of the decay parameter can be numerically unstable. To overcome these limitations, \cite{lu2026mixturesspatialfactoranalyzers} introduce the mixture of spatial factor analyzers (MSFA) with a flexible spatial decay (FSD) covariance structure. The FSD employs I-splines to model the decay curve nonparametrically, yielding a highly adaptable framework and more robust parameter estimation. Consider an observation comprising $q$ non-spatial features measured across a common coordinate system $\mathcal{S}$ with $p$ distinct locations. Let $\mathcal{Y}_i \in \mathbb{R}^{p \times q}$ denote the random matrix formed from the multi-way observations. Within the $g$-th mixture component, the MSFA models $\mathcal{Y}_i$ as
$$ \mathcal{Y}_i = \mathbf{M}_g + \mathcal{U}_{ig} \boldsymbol{\Lambda}_g^{\prime} + \mathcal{E}_{ig},$$
where $\mathbf{M}_g$ represents the location matrix, and $\boldsymbol{\Lambda}_g$ is the column  factor loadings. The latent spatial factor matrix, $\mathcal{U}_{ig} \sim \mathcal{N}_{p \times r}(\mathbf{0}, \boldsymbol{\Sigma}_{1g}, \mathbf{I}_r)$, and the error matrix, $\mathcal{E}_{ig} \sim \mathcal{N}_{p \times q}(\mathbf{0}, \boldsymbol{\Sigma}_{1g}, \boldsymbol{\Psi}_g)$, are assumed to be mutually independent.
Crucially, the spatial covariance matrix $\boldsymbol{\Sigma}_{1g}$ follows the FSD covariance structure
$$ \boldsymbol{\Sigma}_{1g} = \alpha_{1g} \mathbf{J} - \alpha_{2g} \mathbf{D}(\boldsymbol{\beta}_g \mid \mathbf{t},m) + \operatorname{diag}(\boldsymbol{\gamma}_g), $$
where  $\alpha_{1g}, \alpha_{2g} > 0$ are linear spatial parameters, $\mathbf{J}$ is a matrix of ones, and $\mathbf{D}(\cdot)$ contains the I-spline basis functions evaluated over the distance matrix on $\mathcal{S}$, parameterized by the probability simplex coefficients $\boldsymbol{\beta}_g$. The term $\operatorname{diag}(\boldsymbol{\gamma}_g)$ enables modeling of heterogeneous variances across spatial locations. Under this formulation, the conditional distribution of $\mathcal{Y}_i$ is given by
\begin{equation} \label{eq:msfa}
    \mathcal{Y}_i \mid z_{ig} = 1 \sim \mathcal{N}_{p \times q} (\mathbf{M}_g, \alpha_{1g} \mathbf{J} - \alpha_{2g} \mathbf{D}(\boldsymbol{\beta}_g \mid \mathbf{t},m) + \operatorname{diag}(\boldsymbol{\gamma}_g), \boldsymbol{\Lambda}_g\boldsymbol{\Lambda}_g ^{\prime} + \boldsymbol{\Psi}_g).
\end{equation}
Although the constraint mentioned in Section~\ref{sec:identifiability} is not incorporated into \eqref{eq:msfa}, the same consideration applies here.

\subsection{Simulation studies under the finite mixture model context}
We further evaluated the proposed approach in the context of a mixture of matrix normal distributions, where the method performs clustering and imputation simultaneously. With both missing values and component memberships latent, replacing missing entries by conditional means ignores imputation uncertainty and can distort posterior classification probabilities. This may produce overconfident clustering and downward-biased within-component covariance estimates, so the mixture comparison omits the plug-in method and focuses on the other likelihood-based methods. The simulated data are generated from the MMN. The component mean matrices are constructed by adding two component-specific deviations to a common baseline surface. Let $\bar{g} = (G + 1)/2$. For component $g$, the mean matrix is
$$\mathbf{M}_g = \mathbf{M}^{o} + 0.85(g - \bar{g})\mathbf{P}_g + \mathbf{L}_g.$$
The baseline mean $\mathbf{M}^{o}$ is shared by all components and follows a smooth sinusoidal pattern,
$$(\mathbf{M}^{o})_{ij} = \frac{1}{4} \sin \left\{ \frac{i + (j - 1)p}{5} \right\}, \quad i = 1, \dots, p, \quad j = 1, \dots, q.$$
The matrix $\mathbf{P}_g$ gives a smooth component-specific pattern over the row and column directions. With normalized grid points
$$r = -1 + \frac{2(i - 1)}{p - 1}, \quad c = -1 + \frac{2(j - 1)}{q - 1},$$
its entries are
$$(\mathbf{P}_g)_{ij} = \sin \left\{ \frac{(g + 0.5)\pi(r + 1)}{2} \right\} \cos \left\{ \frac{(g + 0.25)\pi(c + 1)}{2} \right\}.$$
The final term $\mathbf{L}_g$ introduces a localized rectangular mean shift, which the $(i,j)$ entry is $(\mathbf{L}_g)_{ij} = 0.55(g - \bar{g})\mathbbm{1}\{(i, j) \in \mathcal{B}_g\}$, where $\mathcal{B}_g$ is a component-specific rectangular subregion and $\mathbbm{1}\{(i, j) \in \mathcal{B}_g\}$ is the corresponding support function. Thus, the components differ through both a smooth global pattern and a local mean shift. When $G = 3$, the second component satisfies $g - \bar{g} = 0$, so its mean matrix is exactly the common baseline $\mathbf{M}^{o}$. The row and column covariance matrices are still generated from an $\operatorname{AR}(1)$ model. For two entries $a=(i_a,j_a)$ and $b=(i_b,j_b)$, the corresponding covariance entries are given by
\begin{equation*}
    (\boldsymbol{\Sigma}_{1g})_{i_ai_b} = \rho_{1g}^{|i_a - i_b|}, \quad \rho_{1g} = 0.35 + 0.04g, \quad \text{and} \quad (\boldsymbol{\Sigma}_{2g})_{j_aj_b} = \rho_{2g}^{|j_a - j_b|}, \quad \rho_{2g} = 0.30 + 0.04g.
\end{equation*}
The scalar variance is $\sigma_g^2 = 0.85 + 0.12g$. The sample size is $N=3000$, and the dimensionality is $15 \times 20$. The comparison here is conducted on the same platform as described in Section~\ref{sec:sim1}. 

In Table~\ref{tab:mixture-15x20}, across both settings, Cholesky closely matches EM in likelihood but provides little computational benefit. The ARI \citep[ARI;][]{hubert1985comparing} and RMSE are omitted from the table because they show little variation across methods within each missingness setting. In contrast, MPEM achieves a substantial speedup while maintaining essentially the same clustering and estimation accuracy with only small likelihood loss. For submatrix missingness, Rect-MPEM is especially efficient, achieving about a $63.0\times$ speedup over the exact EM algorithm while preserving comparable accuracy and yielding a much smaller log-likelihood gap than the nonstructural MPEM. Overall, the experiment shows that the proposed algorithm substantially reduces computational cost while incurring minimal statistical loss for fitting the mixture models.

\begin{table}[t]
\centering
\setlength{\tabcolsep}{3.5pt}
\caption{Mixture experiments with $p=15$, $q=20$, $N=3000$, $G=3$, and nominal missing rate $25\%$. Values are mean (standard deviation) across 30 replications.}
\label{tab:mixture-15x20}
\begin{tabular}{lccc}
\toprule
Method & Time (s) $\downarrow$ & Speedup $\uparrow$ & LL gap / $N$ $\downarrow$ \\
\midrule
\multicolumn{4}{l}{\emph{MCAR}} \\
EM       & 990.3 (320.5)  & 1.00 (0.00)  & 0 \\
Cholesky & 1031.6 (408.4) & 1.01 (0.19)  & $1.67{\times}10^{-8}$ ($3.64{\times}10^{-10}$) \\
MPEM     & 76.9 (12.8)    & 12.63 (2.09) & $3.35{\times}10^{-4}$ ($1.51{\times}10^{-5}$) \\
\addlinespace[2pt]
\multicolumn{4}{l}{\emph{Submatrix}} \\
EM        & 1020.7 (309.9) & 1.00 (0.00)   & 0 \\
Cholesky  & 969.2 (381.9)  & 1.13 (0.25)   & $9.60{\times}10^{-9}$ ($1.77{\times}10^{-9}$) \\
MPEM      & 79.4 (13.8)    & 12.64 (1.80)  & $4.14{\times}10^{-3}$ ($5.09{\times}10^{-4}$) \\
Rect-MPEM & 15.9 (2.3)     & 62.96 (10.56) & $1.21{\times}10^{-4}$ ($2.69{\times}10^{-4}$) \\
\bottomrule
\end{tabular}
\end{table}

\section{Real Application}\label{sec:apps}

The practical efficacy of the proposed imputation method was evaluated by integrating it with the MSFA model and applying the combined approach to the Salinas Valley hyperspectral image dataset \citep{salinas_dataset}. This dataset, collected using the AVIRIS sensor, represents agricultural land cover in California. The complete image provides a spatial resolution of 3.7 meters on a $512 \times 217$ pixel grid and initially contains 224 spectral bands. 
\begin{figure}[ht]
\centering
\includegraphics[width=0.75\linewidth]{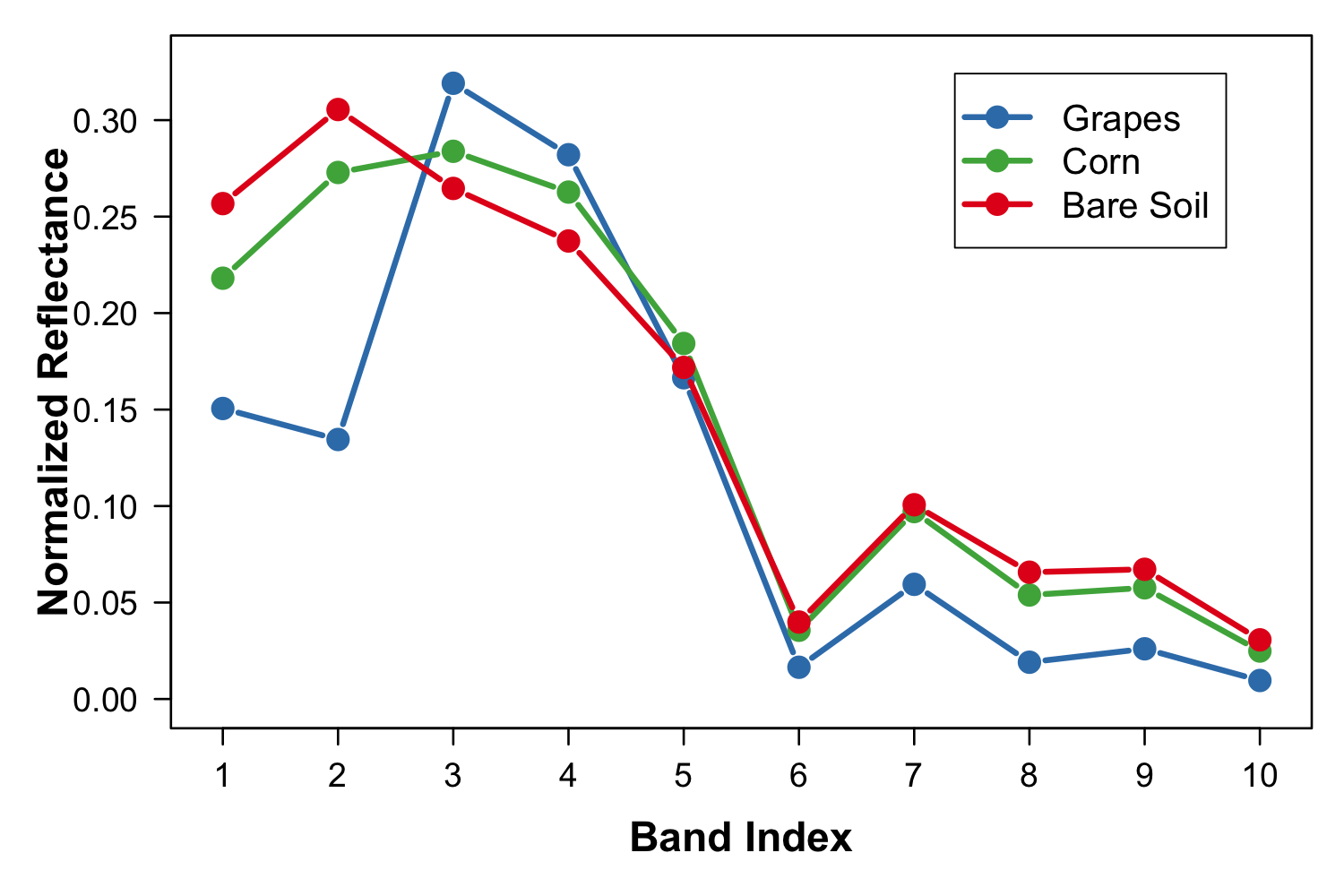}
\caption{Mean spectral profiles of the three classes (Grapes, Corn, and Bare Soil).}
\label{fig: Figure1_Spectra}
\end{figure}
\begin{figure}[ht]
\centering
\includegraphics[width=1\linewidth]{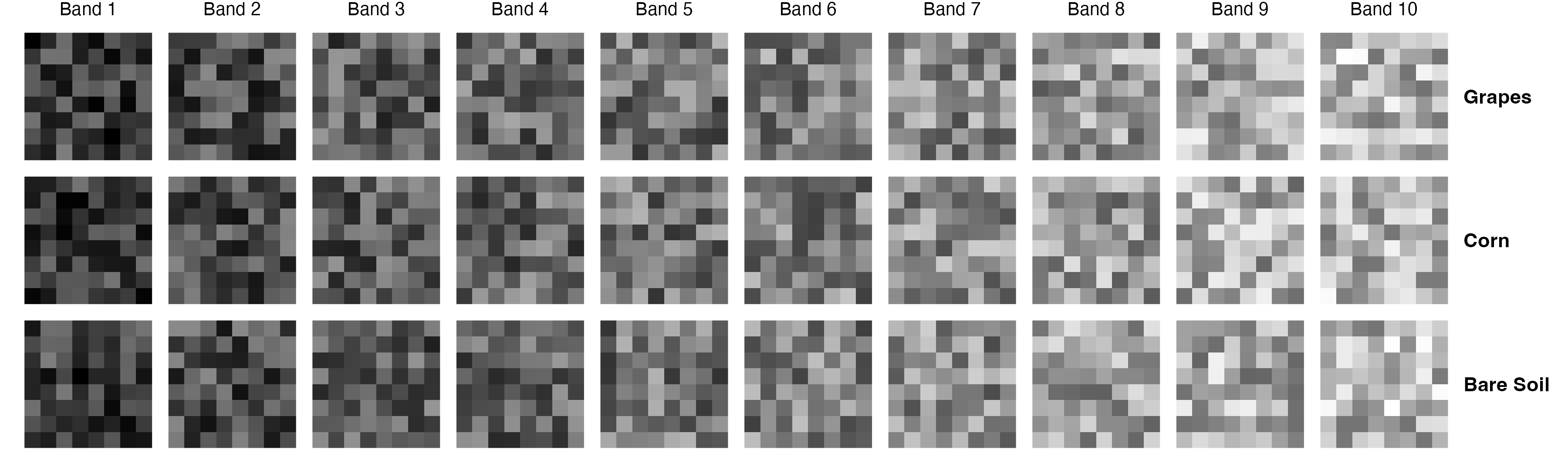}
\caption{Visualization of representative matrix-variate observations of the three classes (Grapes, Corn, and Bare Soil).}
\label{fig: Figure2_Grid_White}
\end{figure}
$25\%$ missing values were introduced using an MCAR and the submatrix patterns. The empirical study focused on the simultaneous imputation and clustering of three distinct land cover types: corn, grapes, and bare soil. To reduce dimensionality, the data were restricted to the most reliable spectral regions by uniformly sampling to retain 10 bands, followed by normalization to the $[0,1]$ interval. Figure~\ref{fig: Figure1_Spectra} shows mean profiles for the retained ten bands. Local spatial dependencies were preserved by dividing the scene into disjoint $8 \times 8$ pixel patches. The class labels were used to construct homogeneous patches and to evaluate clustering, ensuring that every patch belonged entirely to a single class, as illustrated in Figure~\ref{fig: Figure2_Grid_White}. Vectorizing the spatial dimensions of these patches resulted in a final dataset comprising 171 matrix-variate samples of size $64 \times 10$.
\begin{table}[t]
\centering
\setlength{\tabcolsep}{3.5pt}
\caption{Real Salinas missing-data comparison under the constrained spatial mixture model with $\mathbf{M}_g=\mathbf{1}\mu_g^\prime$.}
\label{tab:real3-original-exact-baselines}
\begin{tabular}{lccrcc}
\toprule
Method & RMSE $\downarrow$ & MAE $\downarrow$ & Time (s) $\downarrow$ &
Accuracy $\uparrow$ & BIC $\uparrow$ \\
\midrule
\multicolumn{6}{l}{\emph{MCAR}} \\
EM        & 0.0018 & 0.001 & 67.3 & 0.76 & 404340 \\
MPEM         & 0.0020 & 0.001 & 48.3 & 0.84 & 407633 \\
missRanger   & 0.0068 & 0.004 & 20.7 & 0.85 & 394393 \\
Feature mean & 0.0343 & 0.025 & 0.2  & --   & 306867 \\
\addlinespace[2pt]
\multicolumn{6}{l}{\emph{Structural}} \\
EM        & 0.0047 & 0.002 & 26.6 & --   & 398927 \\
MPEM         & 0.0091 & 0.005 & 2.4  & 0.87 & 407447 \\
Rect-MPEM    & 0.0050 & 0.002 & 1.3  & 0.87 & 408590 \\
missRanger   & 0.0076 & 0.004 & 20.8 & 0.87 & 404327 \\
Feature mean & 0.0276 & 0.020 & 0.2  & $*$  & 367545 \\
\bottomrule
\end{tabular}

\vspace{2pt}
\begin{flushleft}
\footnotesize
\emph{Note:} -- indicates component collapse; $*$ indicates a non-positive-definite covariance update.
\end{flushleft}
\end{table}

The MSFA model with the proposed imputation framework was fitted to the processed data. Various numbers of latent factors were tested, specifically $r \in \{1, \dots, 4\}$, utilizing degree-3 I-splines with 10 knots. Configurations were examined both with and without the constraint $\mathbf{M}_g = \mathbf{1}\boldsymbol{\mu}_g^\prime$, $\boldsymbol{\gamma}_g = \alpha_{3g}\mathbf{1}$. For comparison, the MSFA is fitted via the EM and MPEM. Furthermore, to compare the framework to the general imputation technique, the \texttt{R} package missranger and mean imputation were applied to the flattened feature vectors. Among the MSFA specifications, the model with constrained mean and constrained spatial covariance and $r=3$ yielded an optimal Bayesian Information Criterion \citep[BIC;][]{schwarz1978estimating} of 407,633 and an ARI of 0.8391. Another model, with constrained mean but unconstrained spatial covariance matrices and $r=2$, achieved a higher ARI of 0.943. The comparison results are reported in Table~\ref{tab:real3-original-exact-baselines}. RMSE, MAE, and BIC were computed using the original complete values as ground truth. Accuracy refers to the downstream clustering accuracy after refitting the optimal MSFA to each completed dataset, with labels matched by the best permutation. Accuracy could not be reported when the refitted MSFA either contained fewer than three occupied components (--) or encountered a non-positive-definite covariance update ($*$). A larger BIC indicates better compatibility with the fitted spatial factor analyzer. Under MCAR, the ordinary exact EM gives the smallest imputation error, while the proposed partial EM method is very close in RMSE and MAE but is faster and yields a higher downstream BIC and accuracy. In contrast, missRanger attains competitive classification accuracy but has substantially larger imputation error and lower BIC, indicating weaker agreement with the fitted MSFA. Under submatrix missingness, the proposed structural method is the most effective. It nearly matches the exact EM in imputation accuracy, is over twenty times faster, and gives the largest BIC. The generic partial EM algorithm is very fast but less accurate for this structured pattern, while feature mean imputation is consistently the weakest baseline despite its negligible computational cost. Moreover, for both exact EM and feature mean imputation, the MSFA refitting procedure failed to converge. 

\begin{figure}[ht]
\centering
\includegraphics[width=1\linewidth]{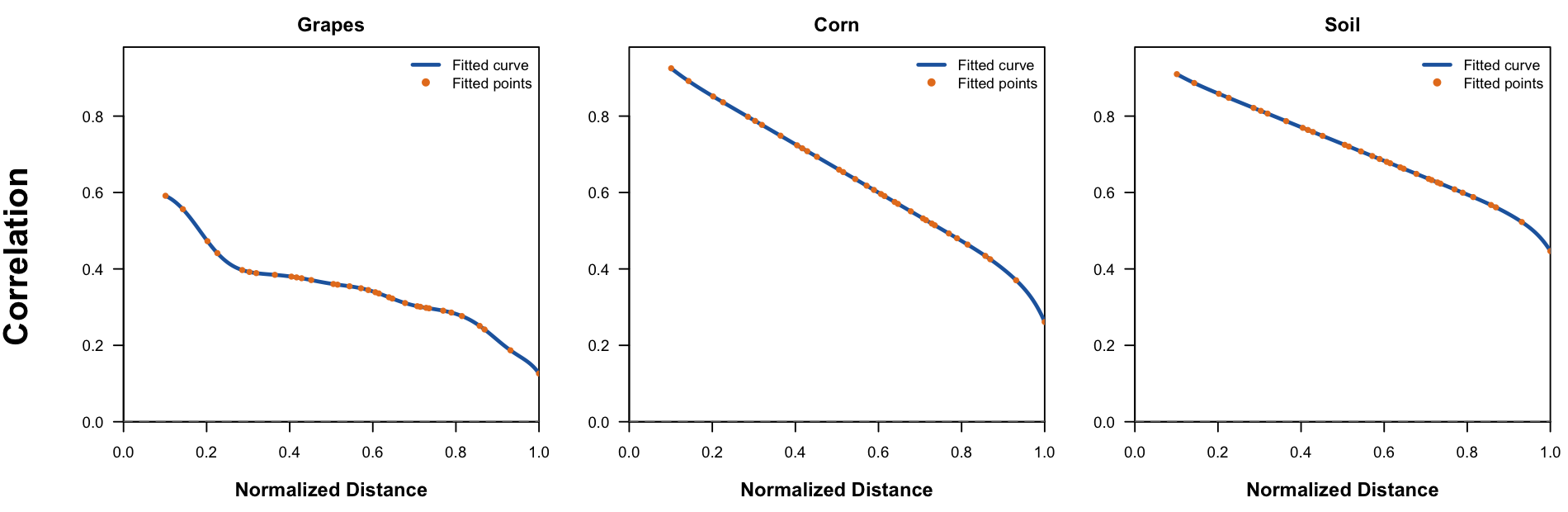}
\caption{Estimated spatial correlation trajectories for the Grapes, Corn, and Bare Soil classes, plotted against normalized distance.}
\label{fig: Real3_curve}
\end{figure}

Figure~\ref{fig: Real3_curve} presents the estimated spatial correlation as a function of normalized distance. The three land-cover categories exhibit inherently distinct spatial profiles, differing substantially in both overall scale and functional form. The clear differentiation among these covariance trajectories and the BIC of the refitted model demonstrate the precision of MPEM's estimation. 

\section{Summary} \label{sec:summary}

In this paper, we proposed an efficient partial EM framework for matrix-variate normal data with missing entries. The main computational burden of the standard EM algorithm comes from the E-step, where the conditional moments of the missing entries must be computed for each missingness pattern. For arbitrary element-wise missingness, this conditioning often disrupts the advantageous Kronecker structure of the matrix normal covariance. To address this issue, coordinate-wise updates are introduced for both the conditional mean and conditional covariance of the missing component. These updates eliminate the need to form the full covariance matrix and to repeatedly invert missing-pattern-specific covariance matrices, while preserving the row-column structure of the matrix normal model.

Structural missingness is also considered, with a focus on cases where the missing entries form a submatrix. In this scenario, the missing-block precision matrix retains a separable Kronecker form. This property enables the conditional covariance update to be performed independently along the row and column directions, further reducing computational cost. 

Simulation studies indicate that the proposed method substantially improves computational efficiency compared with the standard exact EM algorithm, while maintaining nearly identical observed-data likelihood in most scenarios. Although the plug-in method is faster, it may lose accuracy because it ignores the conditional covariance correction. In contrast, the proposed method retains this correction through an efficient approximation. For submatrix missingness, the block version further enhances efficiency over the general version by utilizing the separable structure of the missing block. The real-data application demonstrates that the proposed imputation method can be effectively integrated within a matrix-variate mixture model, enabling simultaneous imputation and clustering while preserving spatial and spectral dependence. Overall, the proposed approach offers a practical and scalable alternative to exact EM for matrix-variate data with both random and structured missingness.

\bibliographystyle{apalike}
\bibliography{EntireLibrary}

\begin{appendices}
\section{Covariance updates for MSFA}
In MSFA, the row covariance matrices are spatially constrained, and the column covariance is constrained by the factor analyzer. Here, the corresponding updates for the parameters in these two component covariance matrices are provided. With the estimates $\widetilde{\boldsymbol{\Sigma}}_{1g}$, the spatial parameters are estimated via the generalized least squares estimator \citep{browne1974generalized}. Let $\mathbf{V}^*_g = (\widetilde{\boldsymbol{\Sigma}}^*_{1g})^{-1}$, where $\widetilde{\boldsymbol{\Sigma}}^*_{1g}$ is the estimate from the last iteration. First, given the current spline coefficients $\hat{\boldsymbol{\beta}}_g$, the linear parameters $\boldsymbol{\alpha}_g = (\alpha_{1g}, \alpha_{2g}, \boldsymbol{\gamma}_g^\prime)^\prime$, where $\boldsymbol{\gamma} = (\alpha_3, \dots, \alpha_{p+2})^\prime$, are updated via
\begin{equation} 
    \hat{\boldsymbol{\alpha}}_g = \left\{\boldsymbol{\Delta}_g^\prime(\mathbf{V}_g^*\otimes \mathbf{V}_g^*)\boldsymbol{\Delta}_g\right\}^{-1}\boldsymbol{\Delta}_g^\prime\operatorname{vec}(\mathbf{V}_g^*\widetilde{\boldsymbol{\Sigma}}_{1g}\mathbf{V}_g^*), 
\end{equation}
where $\widetilde{\boldsymbol{\Sigma}}_{1g}$ is obtained from \eqref{eq:rowcov} in the current iteration, $\boldsymbol{\Delta}_g$ is the design matrix. The structure of $\boldsymbol{\Delta}$ is defined as that the first two columns are $\operatorname{vec}(\mathbf{J})$ and $-\operatorname{vec}(\mathbf{D}(\boldsymbol{\beta}\mid \mathbf{t},m))$, respectively. For $i > 2$, the $i$-th column corresponds to the parameter $\alpha_i$ and is defined as $\operatorname{vec}(\mathbf{E}_{i-2})$, where $\mathbf{E}_{k}$ is a matrix with 1 at the $(k, k)$ entry and 0 elsewhere. Subsequently, when $\hat{\alpha}_{2g} > 0$, using the updated $\hat{\boldsymbol{\alpha}}_g$, we estimate $\boldsymbol{\beta}_g$ by minimizing
\begin{equation*}
    g(\boldsymbol{\beta}) = \frac{1}{2} (\mathbf{r}_g -  \mathscr{I}\boldsymbol{\beta})^\prime(\mathbf{V}_g^*\otimes\mathbf{V}_g^*)(\mathbf{r}_g - \mathscr{I}\boldsymbol{\beta}),
\end{equation*}
 on the probability simplex, where $\mathbf{r}_g = \hat{\alpha}_{2g}^{-1}\operatorname{vec}(\hat{\alpha}_{1g} \mathbf{J} + (\hat{\boldsymbol{\gamma}_g}\mathbf{1}^\prime) \circ \mathbf{I} - \widetilde{\boldsymbol{\Sigma}}_{1g})$, and $\mathscr{I}$ represents the I-spline design matrix.

For the column covariance, define 
\begin{equation*}
    \mathbf{B}_g = \boldsymbol{\Psi}_g^{-1} \boldsymbol{\Lambda}_g 
    \left(\mathbf{I}_r+\boldsymbol{\Lambda}_g^{\prime}\boldsymbol{\Psi}_g^{-1}\boldsymbol{\Lambda}_g\right)^{-1}.
\end{equation*}
Then the update equation for the factor loadings is 
\begin{equation*}
    \hat{\boldsymbol{\Lambda}}_g = \widetilde{\boldsymbol{\Sigma}}_{2g} \mathbf{B}_g \left[ \mathbf{I}_r - \mathbf{B}_g^\prime \boldsymbol{\Lambda}_g + \mathbf{B}_g^\prime \widetilde{\boldsymbol{\Sigma}}_{2g} \mathbf{B}_g \right]^{-1},
\end{equation*}
where $\widetilde{\boldsymbol{\Sigma}}_{2g}$ is obtained from \eqref{eq:colcov}, and 
\begin{equation*}
    \hat{\boldsymbol{\Psi}}_g = \operatorname{diag} \left( \widetilde{\boldsymbol{\Sigma}}_{2g} - \hat{\boldsymbol{\Lambda}}_g \mathbf{B}_g^\prime \widetilde{\boldsymbol{\Sigma}}_{2g} \right).
\end{equation*}
Then, with the estimated parameter, the same calculation as in \eqref{eq:rowcov}, \eqref{eq:colcov} and \eqref{eq:scalarvar} is performed to maintain identifiability.
\end{appendices}

\end{document}